\documentclass{aa}  

\usepackage{graphicx}
\usepackage{multirow}
\usepackage{txfonts}
\bibpunct{(}{)}{;}{a}{}{,} 

\begin{document} 

   \title{Infrared-faint radio sources remain undetected at
   far-infrared wavelengths\thanks{{\it Herschel} is an ESA space observatory
   with science instruments provided by European-led Principal Investigator consortia and with important participation from NASA.}}

   \subtitle{Deep photometric observations using the \textit{Herschel} Space
   Observatory}

   \author{A. Herzog\inst{1,2,3}
	  \and
	  R. P. Norris\inst{3}
      \and
      E. Middelberg\inst{1}
	  \and
	  L. R. Spitler\inst{2,4}
	  \and
	  C. Leipski\inst{5}
	  \and
	  Q. A. Parker\inst{2,4,6}
          }

   \institute{Astronomisches Institut, Ruhr-Universit\"at Bochum, Universit\"atsstr. 150, 44801 Bochum, Germany\\
              \email{herzog@astro.rub.de}
         \and
             Macquarie University, Sydney, NSW 2109, Australia
         \and
            CSIRO Astronomy and Space Science, Marsfield, PO Box 76, Epping, NSW
            1710, Australia
         \and
            Australian Astronomical Observatory, PO Box 915, North Ryde, NSW
            1670, Australia
         \and
            Max-Planck-Institut f\"ur Astronomie, K\"onigsstuhl 17, 69117
            Heidelberg, Germany
         \and
         Department of Physics, Chong Yeut Ming Physics Building, The University
         of Hong Kong, Pokfulam, Hong Kong
         }

   \date{Received 26 November 2014 / Accepted 27 May 2015}

 
  \abstract
   {Showing 1.4\,GHz flux densities in the range of a few to a few tens of mJy,
   infrared-faint radio sources~(IFRS) are a type of galaxy characterised by
   faint or absent near-infrared counterparts and consequently extreme
   radio-to-infrared flux density ratios up to several thousand. Recent studies
   showed that IFRS are radio-loud active galactic nuclei~(AGNs) at redshifts
   $\gtrsim 2$, potentially linked to high-redshift radio galaxies~(HzRGs).}
   {This work explores the far-infrared emission of IFRS, providing
   crucial information on the star forming and AGN activity of IFRS.
   Furthermore, the data will enable examining the putative
   relationship between IFRS and HzRGs and testing whether IFRS are more
   distant or fainter siblings of these massive galaxies.}
   {A sample of six IFRS was observed with the \textit{Herschel} Space
   Observatory between 100\,$\mu$m and 500\,$\mu$m. Using these results, we
   constrained the nature of IFRS by modelling their broad-band spectral energy
   distribution~(SED).
   Furthermore, we set an upper limit on their infrared SED
   and decomposed their emission into contributions from an AGN and
   from star forming activity.}
   {All six observed IFRS were undetected in all five \textit{Herschel}
   far-infrared channels (stacking limits: $\sigma = 0.74$\,mJy at
   100\,$\mu$m, $\sigma = 3.45$\,mJy at 500\,$\mu$m).
   Based on our SED modelling, we ruled out the following objects to explain the photometric characteristics of IFRS:
   (a)~known radio-loud quasars and compact steep-spectrum sources at
   any redshift, (b)~starburst galaxies with and without an AGN and Seyfert
   galaxies at any redshift, even if the templates were modified, and (c)~known
   HzRGs at $z\lesssim 10.5$. We find that the IFRS analysed in this work can
   only be explained by objects that fulfil the selection criteria of HzRGs. More
   precisely, IFRS could be (a)~known HzRGs at very high redshifts ($z\gtrsim
   10.5$), (b)~low-luminosity siblings of HzRGs with additional dust obscuration
   at lower redshifts, (c)~scaled or unscaled versions of Cygnus~A at any
   redshift, and (d)~scaled and dust-obscured radio-loud quasars or compact
   steep spectrum sources. We estimated upper limits on the infrared luminosity,
   the black hole accretion rate, and the star formation rate of IFRS, which
   all agreed with corresponding numbers of HzRGs.}
   {}


   \maketitle
%

\section{Introduction}

A new class of extreme radio galaxies, which are characterised by their
infrared~(IR) faintness and their enormous radio-to-IR flux density ratios, has
recently been found in deep surveys: the class of infrared-faint radio
sources~(IFRS; \citealp{Norris2006}). While various suggestions had been
presented to explain the extreme characteristics of these objects, recent
observations clearly indicate that IFRS are radio-loud~(RL) active galactic
nuclei~(AGNs) at high redshifts.

Infrared-faint radio sources were first discovered in the Chandra Deep Field
South~(CDFS) and the European Large Area Infrared space observatory Survey
South~1 (ELAIS-S1) field of the Australia Telescope Large Area Survey~(ATLAS) by
\citet{Norris2006} and \citet{Middelberg2008ELAIS-S1}, respectively. The radio
maps at 1.4\,GHz provided detections with flux densities in the order of tenths
to tens of mJy, whilst the co-located \textit{Spitzer} Wide-area Infrared
Extragalactic Survey~(SWIRE; \citealp{Lonsdale2003}) showed only a faint or, in
most cases, no IR counterpart at $3.6\,\mu$m with a noise $\sigma =
1\,\mu\mathrm{Jy}$. Later, \citet{Zinn2011} defined the class of IFRS by two
selection criteria:
\begin{enumerate}[(i)]
  \item radio-to-IR flux density ratio
    $S_{1.4\,\textrm{GHz}}/S_{3.6\,\mu\textrm{m}} > 500$, and

  \item 3.6\,$\mu$m flux density $S_{3.6\,\mu\textrm{m}} < 30\,\mu$Jy~.
\end{enumerate}\par
The enormous radio-to-IR flux density ratios imply that IFRS are clear outliers
from the radio-IR correlation. The second criterion implicates a distance
selection, preventing ordinary objects of redshift $\lesssim 1.4$ from being
included in the class of IFRS. In addition to the detections by
\citet{Norris2006} and
\citet{Middelberg2008ELAIS-S1}, IFRS were later found in the  \textit{Spitzer}
extragalactic First Look Survey~(xFLS) field by \citet{GarnAlexander2008}, in
the Cosmological Evolution Survey~(COSMOS) by \citet{Zinn2011}, in the European
Large Area IR space observatory Survey North~1~(ELAIS-N1) by
\citet{Banfield2011}, and in the Lockman Hole field by
\citet{Maini2013submitted}. In total, around 100~IFRS have been found in these
deep fields at a sky density of a few per square degree.\par

Recently, \citet{Collier2014} found 1317~IFRS in shallow all-sky catalogues,
that satisfied both selection criteria given above. They used the Unified Radio
Catalog~(URC; \citealp{Kimball2008}) based on the NRAO VLA Sky Suvey~(NVSS;
\citealp{Condon1998}) and data from the all-sky Wide-Field Infrared Survey
Explorer~(WISE; \citealp{Wright2010}). \citeauthor{Collier2014} replaced
$S_{3.6\,\mu\mathrm{m}}$ by $S_{3.4\,\mu\mathrm{m}}$. All IFRS in that sample
provide an IR counterpart at $3.4\,\mu$m and have radio flux densities above
7.5\,mJy at 1.4\,GHz, some of them exceeding 100\,mJy. By this, the IFRS in this
sample are on average radio-brighter than the IFRS found in the deep fields by
around one order of magnitude. Since the WISE sensitivity is non-uniform,
\citeauthor{Collier2014} could only set a lower limit on the sky density of IFRS
with a radio flux density above 7.5\,mJy of $\sim 0.1\,\mathrm{deg}^{-2}$.\par

First attempts to explain the existence of IFRS included lobes or hotspots of
radio galaxies, obscured star forming galaxies~(SFGs), pulsars or high-redshift
RL galaxies. Later results (for an overview, see \citealp{Collier2014}
and references therein) provided evidence that IFRS are indeed RL AGN at
high redshifts.\par

\citet{Norris2007} and \citet{Middelberg2008IFRS_VLBI} used very long
baseline interferometry~(VLBI) to confirm the presence of active nuclei in
two IFRS. Recently, \citet{Herzog2015a} detected 35 out of 57 IFRS
in VLBI observations, unquestionably confirming that at least a significant
fraction of all IFRS contains AGN.\par

\citet{GarnAlexander2008} and \citet{Huynh2010} present the first spectral
energy distribution~(SED) modelling of IFRS and find that 3C~sources (for
example 3C\,273) can reproduce the observational data when redshifted to
$z\gtrsim 2$. Furthermore, \citeauthor{GarnAlexander2008} and
\citeauthor{Huynh2010} conclude that obscured SFGs cannot explain the
characteristics of IFRS since the radio-to-IR flux density ratios of IFRS
significantly exceed those of SFGs. This finding suggests the presence of an
AGN in IFRS, too.\par

Recently, the first spectroscopic redshifts of IFRS were found, confirming the
suggested high-redshift character of this class of objects. \citet{Collier2014}
find 19~redshifts for their all-sky sample of IFRS, all but one in the range
$2<z<3$. The outlier is expected to be a misidentification or an SFG with an AGN
in its centre. \citet{Herzog2014} measured redshifts of three IFRS in the deep
ATLAS fields based on optical spectroscopy from the Very Large Telescope~(VLT)
and found them to be in the range $1.8<z<2.8$. The IFRS with spectroscopic
redshifts, both from \citeauthor{Collier2014} and \citeauthor{Herzog2014},
lie at the IR-bright boundary of this class of objects with IR~flux densities
between 14\,$\mu$Jy and $30\,\mu$Jy at 3.6\,$\mu$m or 3.4\,$\mu$m.
Both \citeauthor{Collier2014} and \citeauthor{Herzog2014} suggest that the
IFRS with measured spectroscopic redshifts are the closest ones and that
IR-fainter IFRS are at even higher redshifts.\par

Additionally, the results by \citet{Collier2014} and \citet{Herzog2014} finally
disposed of the alternative hypothesis that IFRS are hotspots of spatially
separated radio galaxies. If the IR detections were real counterparts of the
radio detections, the redshift distribution of these objects would be expected
to follow that of Fanaroff Riley type~2 (FRII) galaxies which peaks at $z\sim
1$. In contrast, if the IR detections were spurious associations to the radio
emission, their redshift distribution would be expected to follow that of star
forming galaxies, peaking at $z\sim 0$. Both cases are in clear disagreement
with the observed redshift distribution of IFRS, providing redshifts only in the
regime $1.8\lesssim z\lesssim 3.0$.\par

Based on the redshifts measured in their work, \citet{Herzog2014} present the
first redshift-based SED modelling of IFRS. They find the templates of an
RL AGN (3C\,273) and a compact-steep spectrum~(CSS) source~(3C\,48)
in agreement with all available photometric data, while an ultraluminous IR
galaxy~(ULIRG, Arp\,220) or a Seyfert galaxy~(Mrk\,231) are clearly ruled
out.\par

\citet{Huynh2010} and \citet{Middelberg2011} first pointed out a
potential similarity between IFRS and high-redshift radio galaxies~(HzRGs).
HzRGs are a class of powerful radio galaxies ($L_\mathrm{3\,GHz} >
10^{26}$\,W\,Hz$^{-1}$) at high redshifts ($z \geq 1$).
They are amongst the most massive and most luminous galaxies in the early universe and
are expected to be the progenitors of the most massive galaxies in the local
universe~(e.g.\ \citealp{Seymour2007,deBreuck2010}). Moreover, HzRGs are known
to host powerful AGNs and to show high star formation rates~(SFRs). Therefore,
they are an important tool to study galaxy evolution and the interplay between
AGNs and star forming activity. However, HzRGs are rare objects; only around
100~of these objects are known in the entire sky, located at redshifts $1\leq z
\lesssim 5$.\par

\citeauthor{Huynh2010} note that IFRS have similar mid-IR to radio flux
density ratios to those of HzRGs. \citeauthor{Middelberg2011} studied the radio
properties of IFRS in the ELAIS-S1~field and find steep radio spectra between
2.3\,GHz and 8.4\,GHz with a median of $\alpha = -1.4$\footnote{The spectral
index is defined as $S\propto \nu^\alpha$.} and no index larger than $-0.7$.
This is steeper than the general radio souce population~($\alpha = -0.86$) and
the AGN source population~($\alpha= -0.82$) in that field.
\citeauthor{Middelberg2011} find that the radio spectra are even steeper than
those of the sample of HzRGs by \citet{Seymour2007}, showing a median radio
spectral index $\alpha=-1.02$. HzRGs are the only objects known at significant
redshifts sharing the extreme radio-to-IR flux densities of IFRS in the order of
thousands as shown by \citet{Norris2011}. Furthermore, \citeauthor{Norris2011}
suggest that IFRS might follow the correlation between $3.6\,\mu$m flux
density and redshift found for the sample of HzRGs by \citeauthor{Seymour2007},
similar to the $K-z$ relation for other radio galaxies~\citep{Willott2003}.\par

\citet{Collier2014} and \citet{Herzog2014} tested this hypothesis in the
reachable redshift range between 2 and 3 and find IFRS follow this
correlation. This adds evidence to the suggestion that IR-fainter IFRS are
located at even higher redshift, potentially reaching $z\approx 5$ or 6.
Furthermore, \citeauthor{Herzog2014} show that the radio luminosities of IFRS
are in the same range as for HzRGs, although IFRS lie at the lower bound
of the radio luminosity distribution of HzRGs.\par

Besides these similarities between IFRS and HzRGs, there is one significant
difference. While HzRGs are rare objects of which only around 100 are known in
the entire sky, IFRS are much more abundant. The observed sky density of IFRS in
the ATLAS fields is around $5\,\mathrm{deg}^{-2}$; \citet{Zinn2011} estimate a
survey-independent sky density of $(30.8\pm 15.0)\,\mathrm{deg}^{-2}$.\par

Summarising, there is growing evidence that IFRS are related to HzRGs in the
sense that IFRS are fainter, but much more abundant siblings of HzRGs,
potentially at even higher redshifts. HzRGs are known to be vigorously
forming stars while harbouring AGNs in their centres. Both components dominate
the emission of galaxies in the IR regime. Observations at far-infrared~(FIR)
wavelengths are a key test for the hypothesis that IFRS are related to HzRGs.\par

Here, we analyse the IR and particularly the FIR regime of IFRS
based on \textit{Herschel} observations. Results in this wavelength range will
enable us to further study the suggested link between IFRS and HzRGs.\par

This paper is organised as follows. In Sect.~\ref{observation}, we describe the
sample selection of the observed IFRS and the observations with
\textit{Herschel}. The calibration of the resulting data and the mapping is
presented in Sect.~\ref{reduction}. In Sect.~\ref{photometrystacking}, we
describe the photometry and the consequent stacking analysis. We use the
resulting flux density upper limits in Sect.~\ref{discussion} for two different
approaches to model. First, we model the broad-band SED of IFRS based on
SED templates of known galaxies~(Sect.~\ref{broadbandmodelling}). In a second
approach, we set an upper limit on the IR~SED of IFRS in order to constrain
their IR emission (Sect.~\ref{IRSEDmodelling}). Based on the latter fitting, we estimate the black
hole accretion rate and upper limits for the star formation rate. Finally, in
Sect.~\ref{conclusions}, we present our conclusion. In this paper, we use flat
$\Lambda$CDM cosmological parameters $\Omega_\Lambda = 0.7$, $\Omega_\textrm{M}
= 0.3$, $H_0 = 70$~km~s$^{-1}$~Mpc$^{-1}$ and the calculator by
\citet{Wright2006}.

\section{Sample and observations}
\label{observation}
Six IFRS from the sample compiled by~\citet{Zinn2011} were photometrically
observed with the ESA \textit{Herschel} Space Observatory~\citep{Pilbratt2010},
using the instruments PACS~(Photodetecting Array Camera and Spectrometer;
\citealp{Griffin2010}) and SPIRE~(Spectral and Photometric Imaging Receiver;
\citealp{Poglitsch2010}). The sources were selected to be comparatively bright
in the radio regime, showing 1.4\,GHz flux densities between 7\,mJy and 26\,mJy,
and to provide high radio-to-IR flux density ratios (see
Table~\ref{tab:IFRSsample} also in the following). Since all observed sources
are undetected in the SWIRE data between 3.6\,$\mu$m and 24\,$\mu$m, their
$S_{1.4\,\mathrm{GHz}}/S_{3.6\,\mu\mathrm{m}}$ lower limits based on SWIRE range
between 2300 and 8700. Five observed IFRS are located in the field of ELAIS-S1
and one in the CDFS. No IFRS located in the ELAIS-S1 field has been detected in
the X-ray XMM-Newton survey~(flux limit of $\sim 5.5\times
10^{-16}\,\mathrm{erg\,cm^{-2}\,s^{-1}}$ in the 0.5--2\,keV band;
\citealp{Puccetti2006}) as mentioned by \citet{Zinn2011}. IFRS S703, located in
the CDFS field, has not been covered by the CDFS \textit{Chandra} 4\,Ms
survey~\citep{Xue2011}. No optical counterpart is known for the six observed
IFRS~\citep{Norris2006,Middelberg2008ELAIS-S1}.
\begin{table*}
	\caption{Sample of IFRS observed with \textit{Herschel}.}
	\label{tab:IFRSsample}
 \centering
 \begin{tabular}{c c c c c c c c c c}
 	\hline \hline
	IFRS & RA & $\sigma_\mathrm{RA}$ & Dec & $\sigma_\mathrm{Dec}$ &
	$S_{1.4\,\mathrm{GHz}}$ & $S_{3.6\,\mu\mathrm{m}}$ & $S_{24\,\mu\mathrm{m}}$ &
	$S_{1.4\,\mathrm{GHz}}/S_{3.6\,\mu\mathrm{m}}$ & references \\
	ID & J2000.0 & [arcsec] & J2000.0 & [arcsec] & [mJy] & [$\mu$Jy] & [$\mu$Jy] &
	&
	\\
	\hline S703 & 03:35:31.025 & 0.11 & --27:27:02.20 & 0.12 & $26.1$ & $<2.04$ &
	$<115$ & $>12784$ & (1), (4) \\
	S427 & 00:34:11.592 & 0.01 & --43:58:17.04 & 0.00 & $21.4$ & $1.77\pm 0.54$ &
	$<150$ & $12000$ & (2), (4) \\
	S509 & 00:31:38.633 & 0.01 & --43:52:20.80 & 0.01 & $22.2$ & $<3$ & $<150$ &
	$>7400$ & (2), (3) \\
	S749 & 00:29:05.229 & 0.04 & --43:34:03.94 & 0.04 & $7.0$ & $<3$ & $<150$ &
	$>2337$ & (2), (3) \\
	S798 & 00:39:07.934 & 0.04 & --43:32:05.83 & 0.03 & $7.8$ & $<3$ & $<150$ &
	$>2597$ & (2), (3) \\
	S973 & 00:38:43.489 & 0.04 & --43:19:26.94 & 0.04 & $9.1$ & $<2.70$ & $<150$ &
	$>3385$ & (2), (4)
	\\
	\hline
	\end{tabular}
\tablefoot{Infrared-faint radio source S703 is located in CDFS while the other five
	sources are located in the ELAIS-S1 field. All six IFRS are undetected at
	3.6\,$\mu$m in the SWIRE survey ($3\sigma \sim 3\,\mu$Jy) as presented by
	\citet{Zinn2011}. \citet{Maini2013submitted} analysed the deeper SERVS data for
	three IFRS observed with Herschel and find one week counterpart. All upper
	limits represent $3\sigma$. S973 consists of two radio components. Here, we used the position of the strongest component
	as the position of the source, in contrast to
	\citeauthor{Middelberg2008ELAIS-S1} who used the centre between both
	components. Positions, position uncertainties, radio and 24\,$\mu$m
	flux densities were taken from the first reference listed in each row,
	$S_{3.6\,\mu\mathrm{m}}$ and $S_{1.4\,\mathrm{GHz}}/S_{3.6\,\mu\mathrm{m}}$
	from the second reference.}
	\tablebib{(1)~\citet{Norris2006};
	(2)~\citet{Middelberg2008ELAIS-S1}; (3)~\citet{Zinn2011};
	(4)~\citet{Maini2013submitted}.}
\end{table*}
By these characteristics, the six observed sources are prototypical
for the class of IFRS. We note that \citet{Maini2013submitted} find a
$3\sigma$ counterpart at 3.6\,$\mu$m based in the \textit{Spitzer}
Extragalactic Representative Volume Survey~(SERVS; \citealp{Mauduit2012})
for IFRS~S427 after the observations with \textit{Herschel} were
carried out.\par

\begin{figure}
	\centering
		\includegraphics[width=\hsize]{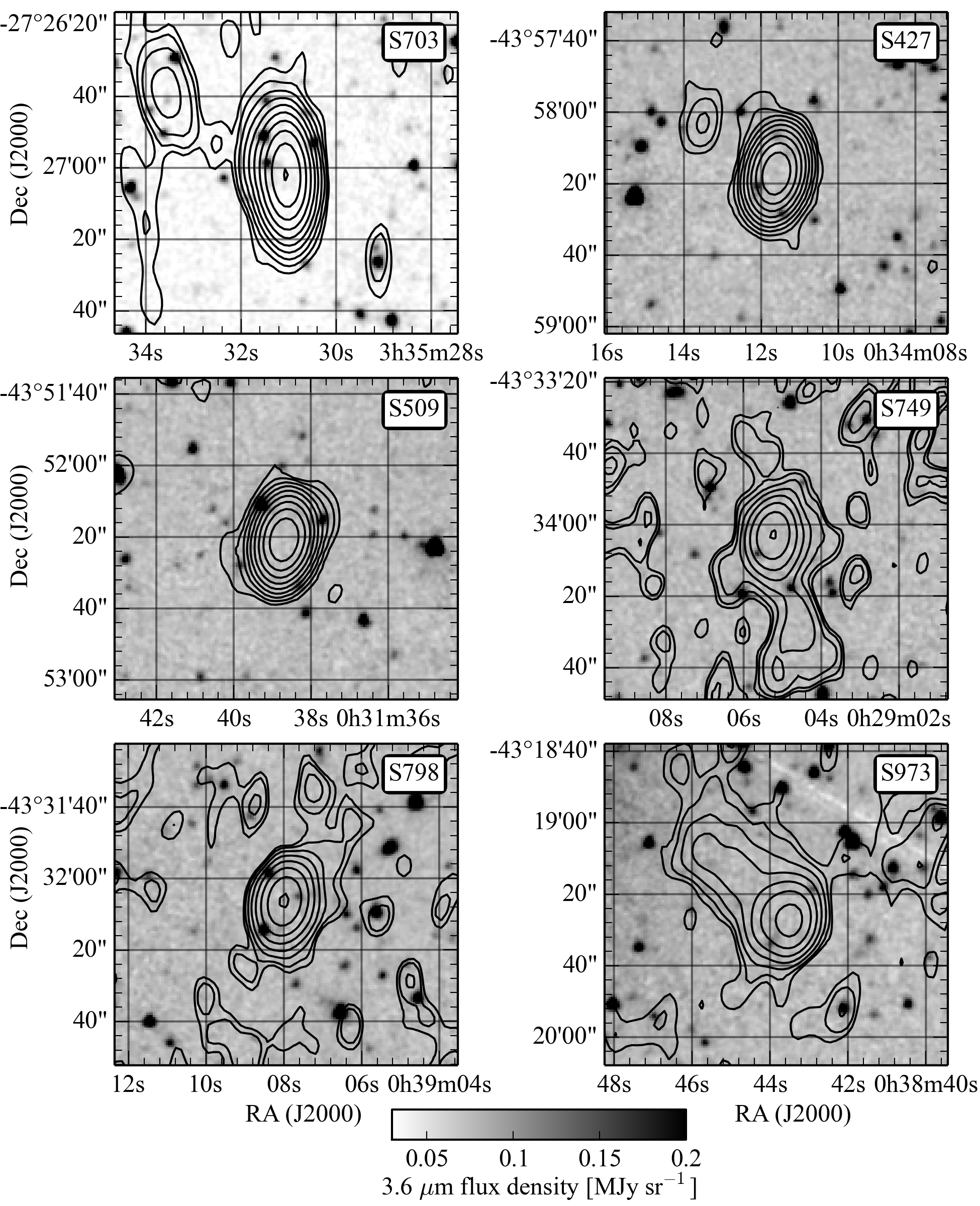}
		\caption{SWIRE 3.6\,$\mu$m maps~(greyscale; \citealp{Lonsdale2003}) of all six
		IFRS observed with \textit{Herschel} overlayed by the 1.4\,GHz radio contours
		(ATLAS data release~3; $\sigma \approx
		20\,\mu\mathrm{Jy\,beam}^{-1}$; \citealp{Franzen2015submitted}).
		Shown are the contours at $2\sigma$, $4\sigma$, $8\sigma$, $16\sigma$, etc.
		Top down and from left to right: S703, S427, S509, S749, S798, and S973.}
	\label{fig:SWIRE_allIFRS}
\end{figure}

Figure~\ref{fig:SWIRE_allIFRS} shows the SWIRE 3.6\,$\mu$m maps of all IFRS
observed with \textit{Herschel}, overlayed by the 1.4\,GHz radio contours.
IFRS S427, S509, and S798 are point-like sources and show almost no extended
structure. For S703, in contrast, it might be suggested because of the
additional radio component at around RA\,03h35m33.5s Dec\,$-27^\circ26'40"$
and the bridge to the IFRS itself that the IFRS is part of a double lobe
radio source with the IFRS and the additional component as radio lobes and the
host galaxy in between those two components. However, the additional radio
component shows a weak 3.6\,$\mu$m counterpart, excluding this component as a
radio lobe. Therefore, S703 can be considered as a proper IFRS.\par

Infrared-faint radio source S973 shows an extended structure and has been listed
as two components (RA\,00h38m44.723m Dec\,$-43^\circ 19'14.58''$ and
RA\,00h38m43.489s Dec\,$-43^\circ 19'26.94''$) by
\citet{Middelberg2008ELAIS-S1}, with the source position in between the two
components. Here, however, we considered the position of the
second, radio-brighter component to be the centre of the host with the weaker
component more likely to be a jet. However, the choice of host position does
not affect our overall conclusion. The position of the brighter component is
listed in Table~\ref{tab:IFRSsample}.\par

Infrared-faint radio source S749 shows an extended structure in the radio map,
too. However, this source is located close to the edge of the ATLAS ELAIS-S1
field and, therefore, the noise is higher than for the other IFRS discussed in this
work. In the scheduling of the \textit{Herschel} observations, a slightly wrong position
of IFRS~S749 had been used. Thus, this source is not located in the centre of
the \textit{Herschel} maps but closer to the edge, resulting in a higher noise compared
to the other sources, particularly at shorter wavelengths. Therefore, we
will exclude this source from our stacking analysis in Sect.~\ref{photometrystacking}.\par

The PACS observations of these six objects were carried out in July and
December~2011 (observation IDs 1342224373, 1342224374, 1342233615, 1342233616,
1342233617, 1342233618, 1342233619, 1342233620, 1342233621, 1342233622,
1342233623, 1342233624). Using the mode \texttt{mini-scan map}, each observation
was divided into two parts, arising from two different scan angles. Both
scanning parts were centred on the source, observing simultaneously at
100\,$\mu$m and 160~$\mu$m with a total on-source time of 22.5~minutes for each
source.\par

The SPIRE observations were carried out in December~2011 and January~2012
(observation IDs 1342234729, 1342234730, 1342234731, 1342234732, 1342234733,
1342238290). Using the observing mode \texttt{SpirePhotoSmallScan}, the sources
were observed simultaneously at 250\,$\mu$m, 350\,$\mu$m and 500\,$\mu$m for
12~minutes each.\par

We recognise that, since these observations were made and this analysis
was performed, additional Herschel data on IFRS have been released into the
public domain by the \textit{Herschel} Multi-Tiered Extragalactic Survey~(HerMES;
\citealp{Oliver2012}), the PACS Evolutionary Probe~(PEP; \citealp{Lutz2011}),
and the Cosmic Assembly Near-infrared Deep Extragalactic Legacy
Survey~(CANDELS; \citealp{Grogin2011}). These data have not been used for the
study presented in this work, but will be discussed in a future paper.\par

\section{Data calibration and mapping}
\label{reduction}
\begin{figure}
	\centering
		\includegraphics[width=\hsize]{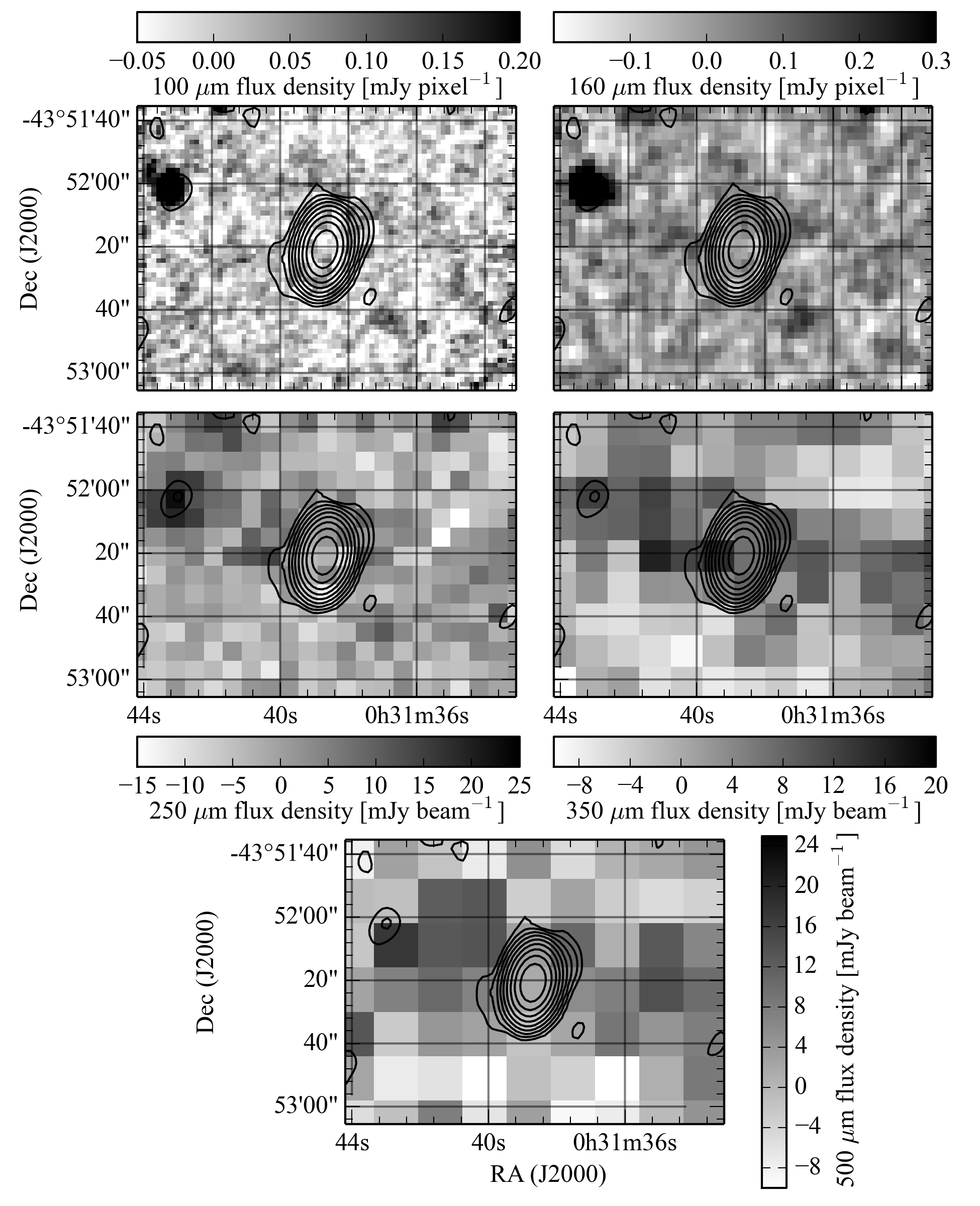}
		\caption{Final \textit{Herschel} maps~(greyscale) of IFRS~S509 overlayed by
		the 1.4\,GHz radio contours (ATLAS data release~3; $\sigma \approx
		20\,\mu\mathrm{Jy\,beam}^{-1}$; \citealp{Franzen2015submitted}).
		Shown are the contours at $2\sigma$, $4\sigma$, $8\sigma$, $16\sigma$, etc.
		Top down and from left to right: 100\,$\mu$m, 160\,$\mu$m (both PACS),
		250\,$\mu$m, 350\,$\mu$m, and 500\,$\mu$m (SPIRE) maps. None of the five maps
		provides a detection of S509.}
	\label{fig:mapsS509}
\end{figure}

The data sets were calibrated using the \textit{Herschel} Interactive Processing
Environment~(HIPE; \citealp{Ott2010}, version~12.1.0). We followed the steps
presented in the PACS photometer pipeline for deep survey maps provided within
HIPE to process the PACS data sets from level~0 to level~2. During the
processing to level~1, a mask was created and applied in the task
\texttt{highpass filter} to prevent artefacts in this median-subtraction process
arising from nearby bright sources. The position of the IFRS has also
been masked. We obtained the best results using a highpass filter radius of
$15\,\arcsec$. The final maps were built using the task \texttt{photProject},
setting the parameter \texttt{pixfrac} to 0.5, both for 100\,$\mu$m and
160\,$\mu$m. Output pixel sizes of 1.5\arcsec~(for 100\,$\mu$m) and
2.1\arcsec~(for 160\,$\mu$m), respectively, were found to provide the best maps.
Both quantities are parameters of the task \texttt{photProject} which is based
on the Drizzle method~\citep{FruchterHook2002}. Finally, we matched the two maps
of each object, obtained from the two different scan angles, using the task
\texttt{mosaic}.
\begin{table*}
	\caption{Characteristics of the instruments and parameters of the data
	reduction and photometry, used in HIPE for the different channels.}
	\label{tab:redphot}
 \centering
 \begin{tabular}{c c c c}
 	\hline \hline
	Band & FWHM & Pixel size & Exposure time \\
	 & [arcsec$^2$] & [arcsec pixel$^{-1}$] & [s] \\ \hline
	PACS 100\,$\mu$m & 6.69~$\times$~6.89 & 1.5 & 1354 \\
	PACS 160\,$\mu$m & 10.65~$\times$~12.13 & 2.1 & 1354 \\
	SPIRE 250\,$\mu$m & 17.6~$\times$~17.6 & 6 & 721 \\
	SPIRE 350\,$\mu$m & 23.9~$\times$~23.9 & 10 & 721 \\
	SPIRE 500\,$\mu$m & 35.2~$\times$~35.2 & 14 & 721 \\
	\hline
	\end{tabular}
	\tablefoot{The numbers of the FWHM are taken from
	``PACS Observer's Manual'' version~2.5.1 and from ``The SPIRE Handbook'' version~2.5, respectively.}
\end{table*}
The final PACS~maps have a size of around $3.5\arcmin\times 6.5\arcmin$. The
central region with a diameter of around 50\arcsec provides the highest and an
almost uniform sensitivity. Figure~\ref{fig:mapsS509} (upper subplots)
shows the final PACS maps of IFRS~S509.\par

Correspondingly, for the SPIRE data sets, we followed the steps in the
appropriate HIPE standard pipeline---the ``Photometer Small Map user
pipeline''---to calibrate the data. The maps were built using the task
\texttt{Destriper}, an iterative algorithm to remove the baseline signal in the
timeline data. In this task, the implemented deglitcher was used. We
used the algorithm \texttt{naive mapping} implemented in HIPE to build the final
maps, projecting the full power seen by each bolometer timeline step to the
closest sky map pixel with the pixel sizes given in Table~\ref{tab:redphot}. The
usable area of the final SPIRE maps is around $8\arcmin\times 10\arcmin$.
Figure~\ref{fig:mapsS509} (lower three subplots) shows the resulting
SPIRE maps of IFRS~S509.

\section{Photometry and stacking}
\label{photometrystacking}

\subsection{Photometry}
\label{photometry}
We used the HIPE task \texttt{AnnularSky\-Aperture\-Photometry} for photometry
on the PACS maps, performing aperture photometry on a chosen target based on a
circular aperture. The related task \texttt{PhotApertureCorrectionPointSource}
corrected the measured flux density for the finite size of the applied aperture,
yielding the requested flux density of the nominal source. We used
aperture radii of 5.6\arcsec and 10.5\arcsec for the PACS maps at 100\,$\mu$m
and 160\,$\mu$m, respectively. The background was estimated und subtracted based
on a ring between 20\arcsec and 25\arcsec, and 24\arcsec and 28\arcsec,
respectively. The flux density uncertainties were obtained from 100~similar
apertures randomly placed on the map where the coverage was at least 75\% of the
maximum coverage. We fit a Gaussian to the histogram of these background flux
densities, providing the Gaussian width~$\sigma$. We repeated this procedure ten
times and took the median of these Gaussian widths as the uncertainty of the
measured flux density. For S749, where the IFRS was not in the centre
of the \textit{Herschel} maps as discussed in Sect.~\ref{observation}, we
placed 100~similar apertures in a field with a maximum distance to the source of five
times the full width at half maximum~(FWHM) of a point source to estimate the
flux uncertainty.\par

Our six nominal objects were not detected at 100\,$\mu$m or at
160\,$\mu$m. Table~\ref{tab:FluxResults} summarises the resulting point source
flux density uncertainties~$\sigma$ for the six observed IFRS.\par

Photometry on the SPIRE maps was carried out using the HIPE task
\texttt{sourceExtractorSussextractor} based on the SUSSEXtractor
algorithm~\citep{SavageOliver2007}. Adapting this task to the maps, sources were
extracted, providing data about brightnesses, positions, and related errors of
the sources as well as a cleaned map. The extracted sources were checked for
their angular distance to the known position of IFRS ($<0.5$\,FWHM). We repeated
this procedure as a second level source extraction on the cleaned map.
Again, we checked the extracted sources for their distance to the IFRS'
positions. By this, we were able to extract sources which had been overseen in
the first extraction step because of bright and close by sources in the map.
These sources had been eliminated in the cleaned map.\par

For five out of the six IFRS, the source extraction did not provide any
FIR component in any SPIRE channel, either by the source extraction performed on
the map itself or on the cleaned map. For S973, we found $3\sigma$ detections in
the cleaned maps at 350\,$\mu$m and 500\,$\mu$m. The positions of these two FIR
counterparts were in agreement with each other and were 12\arcsec distant from
the IFRS radio position, which is only slightly below our distance criterion.
However, we found that these FIR detections were very close to one source which
was detected with a flux density of 6.1\,$\mu$Jy at 3.6\,$\mu$m in the SERVS
survey. Because of those three reasons---weak flux, relatively large
distance to the IFRS position, and overlap with a SERVS source---we suggest that
these FIR counterparts are associated with the SERVS source and not with our
SERVS-undetected IFRS.

We summarise that none of the six observed IFRS was detected in any SPIRE
channel. The flux density uncertainties were obtained from the cleaned map by
fitting a Gaussian to the pixel values within a square of the size of eight times the FWHM
of a point source (see numbers in Table~\ref{tab:redphot}).
\begin{table*}
\caption{Resulting point source flux density uncertainties for the six IFRS at
the five observed wavelengths. The last column shows the respective
uncertainties in the stacked maps, resulting from a median stack of five
individual maps at each wavelength.}
\label{tab:FluxResults}
 \centering
 \begin{tabular}{c c c c c c c c}
 	\hline \hline
 	 Band & $\sigma_\mathrm{S703}$ & $\sigma_\mathrm{S427}$ &
 	 $\sigma_\mathrm{S509}$ & $\sigma_\mathrm{S749}$ & $\sigma_\mathrm{S798}$ &
 	 $\sigma_\mathrm{S973}$ & $\sigma_\mathrm{stacked}$ \\
	 & [mJy] & [mJy] & [mJy] & [mJy] & [mJy] & [mJy] & [mJy] \\ \hline
 	PACS 100\,$\mu$m & 1.78 & 1.40 & 1.83 & 2.17 & 1.51 & 1.43 & 0.76 \\
 	PACS 160\,$\mu$m & 4.02 & 3.56 & 2.43 & 6.09 & 2.84 & 2.77 & 1.66 \\
 	SPIRE 250\,$\mu$m & 5.03 & 4.20 & 4.51 & 4.43 & 4.94 & 4.99 & 2.68 \\
 	SPIRE 350\,$\mu$m & 4.61 & 3.81 & 4.42 & 4.59 & 3.87 & 5.71 & 2.52 \\
 	SPIRE 500\,$\mu$m & 5.50 & 4.80 & 4.85 & 6.24 & 5.36 & 7.23 & 3.53\\
	\hline
	\end{tabular}
\end{table*}
The SPIRE flux density uncertainties at the positions of the IFRS are
summarised in Table~\ref{tab:FluxResults}. Our mean noise
is lower than the overall confusion noise of 5.8\,mJy, 6.3\,mJy,
and 6.8\,mJy at 250\,$\mu$m, 350\,$\mu$m, and 500\,$\mu$m,
respectively~\citep{Nguyen2010}. However, \citeauthor{Nguyen2010} also showed
that the residual confusion noise after removing bright sources is lower. Our
noises are in agreement with these numbers.\par

\subsection{Stacking of \textit{Herschel} maps}
\label{stackingHerschel}
\begin{figure}
	\centering
		\includegraphics[width=\hsize]{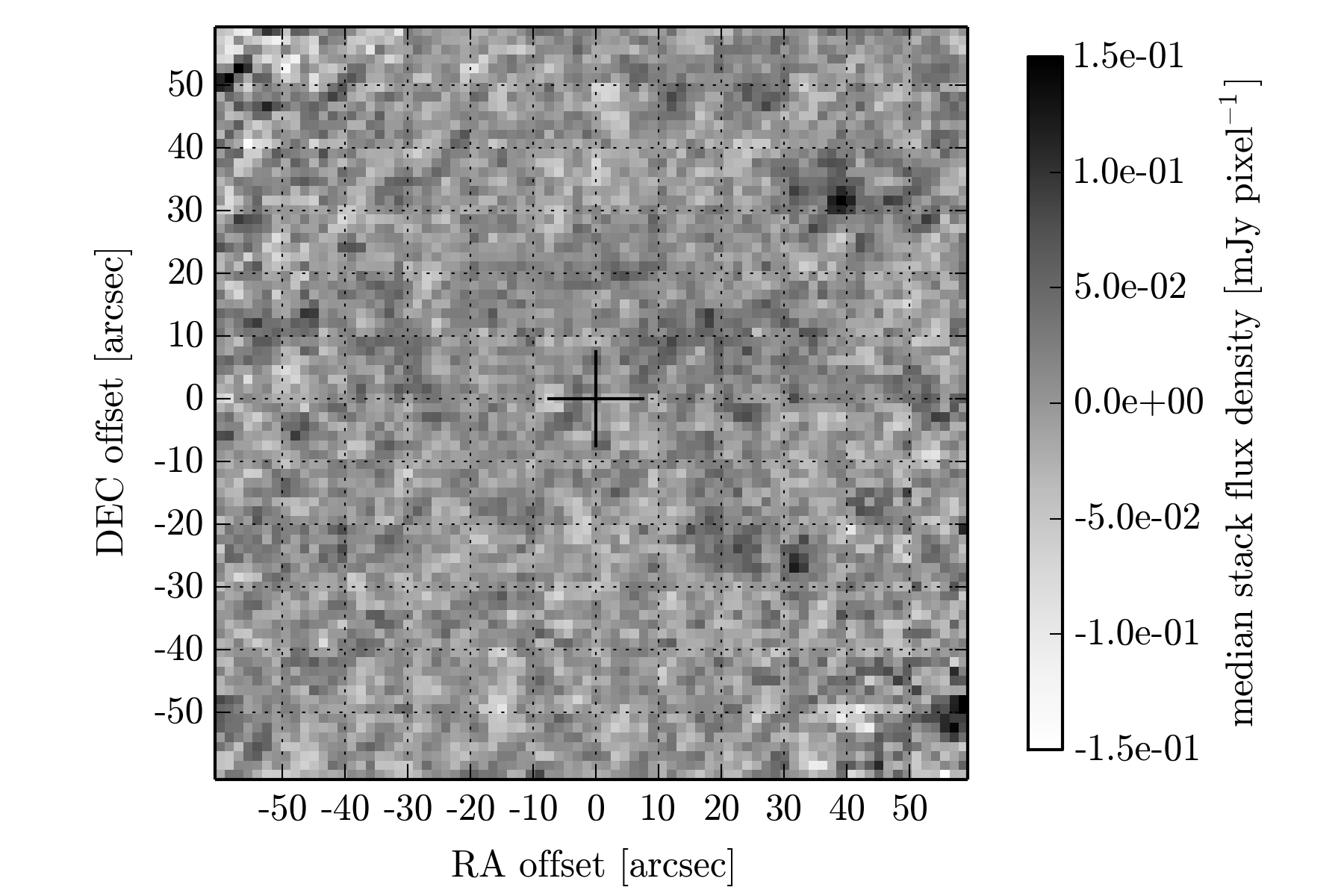}
		\caption{Resulting stacking map obtained from a median stack of the five
		IFRS at 100\,$\mu$m. IFRS S749 was excluded because of the higher noise as
		discussed in Sect.~\ref{observation}. The individual maps were centred on the
		position of the IFRS which is marked by a cross in the centre of the map. No
		detection has been found in the stacking map.}
	\label{fig:Stacking}
\end{figure}

Since no counterpart of an IFRS was detected in the \textit{Herschel} maps, we
performed a stacking analysis to search for a potentially weak counterpart
slightly below the detection limit. The positional uncertainties for our sources
are at least one order of magnitude lower than the pixel size of the
\textit{Herschel} maps (see Tables~\ref{tab:IFRSsample} and \ref{tab:redphot}).
For each observed wavelength, we stacked the maps centred on the known position
of the IFRS. However, we excluded IFRS~S749 from the stacking analysis because
of the higher noise in the FIR~maps (see discussion in Sect.~\ref{observation}).
The stacking maps are the results of median stacking the Herschel maps for all
the five sources with proper centring and signal-to-noise ratio at each
wavelength. In case of SPIRE maps, we stacked the cleaned maps,
resulting from the first iteration of source extraction described above. The
stacking map at 100\,$\mu$m is shown in Fig.~\ref{fig:Stacking} and did not
provide a detection. Corresponding stacking maps at 160\,$\mu$m, 250\,$\mu$m,
350\,$\mu$m, and 500\,$\mu$m, respectively, were similar and did not provide a
detection either. We performed photometry on the stacked maps in the same way as
described in Sect.~\ref{photometry} for the individual maps. The resulting flux
density uncertainties are summarised in the last column in
Table~\ref{tab:FluxResults}.

\subsection{Stacking in the near- and mid-infrared regime}
\label{stackingIR}
Since the six observed IFRS were also undetected in the SWIRE maps at
24\,$\mu$m, we performed a stacking analysis at these wavelengths to obtain mean
properties of our IFRS population in the following analysis.\par

At 3.6\,$\mu$m, all six IFRS were covered by SWIRE, but only four by the
deeper SERVS survey. We median stacked the 3.6\,$\mu$m maps at the positions of
the six sources. We carried out aperture photometry using an annulus of
1.9\arcsec in radius and a background ring between 3.8\arcsec and 7.6\arcsec and
applied aperture corrections. No source was found in the stacked map.
The uncertainty was obtained based on 225~randomly placed apertures and
background rings of same size and iteratively removing those apertures resulting
in flux densities exceeding $3\sigma$. We found an uncertainty of
$\sigma_\mathrm{3.6\,\mu m} = 0.397\,\mu$Jy in the stacked map.\par

At 24\,$\mu$m, all six IFRS were covered by SWIRE. We median stacked the
six individual maps and carried out aperture photometry using an annulus of
3.5\arcsec in radius and a background ring between 6.0\arcsec and 8.0\arcsec
and applying aperture corrections. No source was found in the stacked
map either. We obtained the uncertainty in the same was as described
for the stacked map at 3.6\,$\mu$m and found it to be $\sigma_\mathrm{24\,\mu m}
= 31.3\,\mu$Jy.\par

\section{Modelling and analysis}
\label{discussion}
The \textit{Herschel} observations between 100\,$\mu$m and 500\,$\mu$m
of the six IFRS presented in this work did not provide a detection. A stacking analysis
did not show an FIR counterpart below the detection limit either. However, the
flux density upper limits measured in Sect.~\ref{photometrystacking} put
constraints on the SED of IFRS.\par

The six IFRS observed with \textit{Herschel} were selected to show high
radio-to-IR flux density ratios ($>2000$; see Sect.~\ref{observation}). Five of
these IFRS are undetected in the near-IR regime while SERVS provides a
$3.2\sigma$ couterpart at 3.6\,$\mu$m for one IFRS as reported by
\citet{Maini2013submitted}. This counterpart was unknown at the time when the
observations with \textit{Herschel} were carried out. Except for this single
near-IR detection, these six IFRS are solely detected in the radio regime. Thus,
no redshift is known for these sources. By these characteristics, the six IFRS
observed with \textit{Herschel} are among the most extreme objects in the class
of IFRS.\par

All IFRS with known spectroscopic redshifts, presented by \citet{Collier2014}
and \citet{Herzog2014}, are faintly detected in the near-IR regime.
Their redshifts, $1.8\lesssim z \lesssim 3$, are in agreement with the
suggestion by \citet{Norris2011} that IFRS follow the correlation between
3.6\,$\mu$m flux density and redshift found for HzRGs by \citet{Seymour2007}.
Assuming that this correlation holds at lower IR~flux densities, as it does for
HzRGs, the IFRS observed with \textit{Herschel} would be placed at redshifts
of~$z \gtrsim 4$ or higher because of their 3.6\,$\mu$m faintness.\par

We emphasise that no redshift above $z\gtrsim 3$ has been measured
for an IFRS, presumably because those are too faint for spectroscopic
observations. Although we speculate that the IFRS discussed here are at
$z\gtrsim 4$, we will consider the broad redshift range $1 \leq z \leq 12$.\par

In the following, we tried to limit possible explanations for the
phenomenon of IFRS, performing a broad-band SED modelling based on SED templates
of known galaxies~(Sect.~\ref{broadbandmodelling}). We started with the simplest
approach and shifted these templates in a broad redshift range, testing them
against the photometric data of IFRS (Sect.~\ref{differentredshift}).
Subsequently, we broadened the parameter space by modifying our broad-band SED
templates. We scaled the templates in luminosity (Sect.~\ref{scaling}), added
extinction (Sect.~\ref{extinction}), and finally modified them simultaneously in
luminosity and extinction (Sect.~\ref{scalingextinction}). In each approach, we tested their
compatibility with the photometric constraints of IFRS . Finally, we constrained
the IR~SED of IFRS based on decomposing their IR SED into an AGN and a
starburst~(SB) component (Sect.~\ref{IRSEDmodelling}), both putatively
contributing to the total emission of this peculiar class of objects.

\subsection{Broad-band SED modelling}
\label{broadbandmodelling}

In order to model the broad-band SED of IFRS, we used photometric data
of our objects in all available wavelength regimes. The IFRS observed with
\textit{Herschel} provided 1.4\,GHz flux densities between 7\,mJy and 26\,mJy.
In the modelling, we used a median 1.4\,GHz flux density of 15\,mJy, but also
discuss the outcome for higher and lower radio flux densities. Furthermore, we used
available constraints at 3.6\,$\mu$m and 24\,$\mu$m. At 3.6\,$\mu$m, we used a
median $3\sigma$ flux density upper limit of 1.19\,$\mu$Jy as discussed in
Sect.~\ref{stackingIR}. In the mid-IR at 24\,$\mu$m, we used a $3\sigma$ flux
density upper limit of 94\,$\mu$Jy based on the stacking described in
Sect.~\ref{stackingIR}. Furthermore, we made use of the far-IR flux density
upper limits at five wavelengths between 100\,$\mu$m and $500\,\mu$m measured in
this work. More precisely, we used $3\sigma$ flux density upper limits based on
the uncertainties in the median stacked maps presented in
Sect.~\ref{stackingHerschel} and summarised in the last column in
Table~\ref{tab:FluxResults}.\par

Based on photometric data and redshifts from the NASA/IPAC Extragalactic
Database~(NED), we built SED templates for different objects by connecting the
data points and smoothing the template where appropriate. We used templates of
the spiderweb galaxy~(an HzRG, also known as MRC\,1138-262), the local
radio galaxy Cygnus~A (also known as 3C\,405), the CSS source~3C\,48, the RL
quasar~3C\,273, the local ULIRG Arp\,220, the Seyfert galaxy Mrk\,231, the local
SB galaxy M82, an RL hyper-luminous infrared galaxy~(HyLIRG; IRAS~F15307+3252),
IRAS~F00183-7111 (referred to as F00183; this object is a ULIRG, showing
contribution of an RL~AGN combined with significant SB activity), and the
quiescent elliptical brightest cluster galaxy~NGC\,1316. We added
photometric IR data between 100\,$\mu$m and 870\,$\mu$m from \citet{Seymour2012} to the
template of the spiderweb galaxy since the IR coverage of this template is poor
in NED and this wavelength regime is crucial for our analysis. All photometric
data points from \citeauthor{Seymour2012} are at least $4\sigma$
detections.\par

\subsubsection{Shifting broad-band SED templates to various redshifts}
\label{differentredshift}

As a simplest approach, we tested whether any SED template of a known
galaxy is in agreement with the available photometric data of IFRS when these
templates were shifted to different redshifts, keeping constant rest-frame
luminosity and scaling flux densities using a flat $\Lambda$CDM cosmology ($\Omega_\Lambda = 0.7$, $\Omega_\mathrm{M} = 0.3$, $H_0 =
70\,\mathrm{km\,s^{-1}\,Mpc^{-1}}$).
\begin{figure*}
	\centering
		\includegraphics[width=8.5cm]{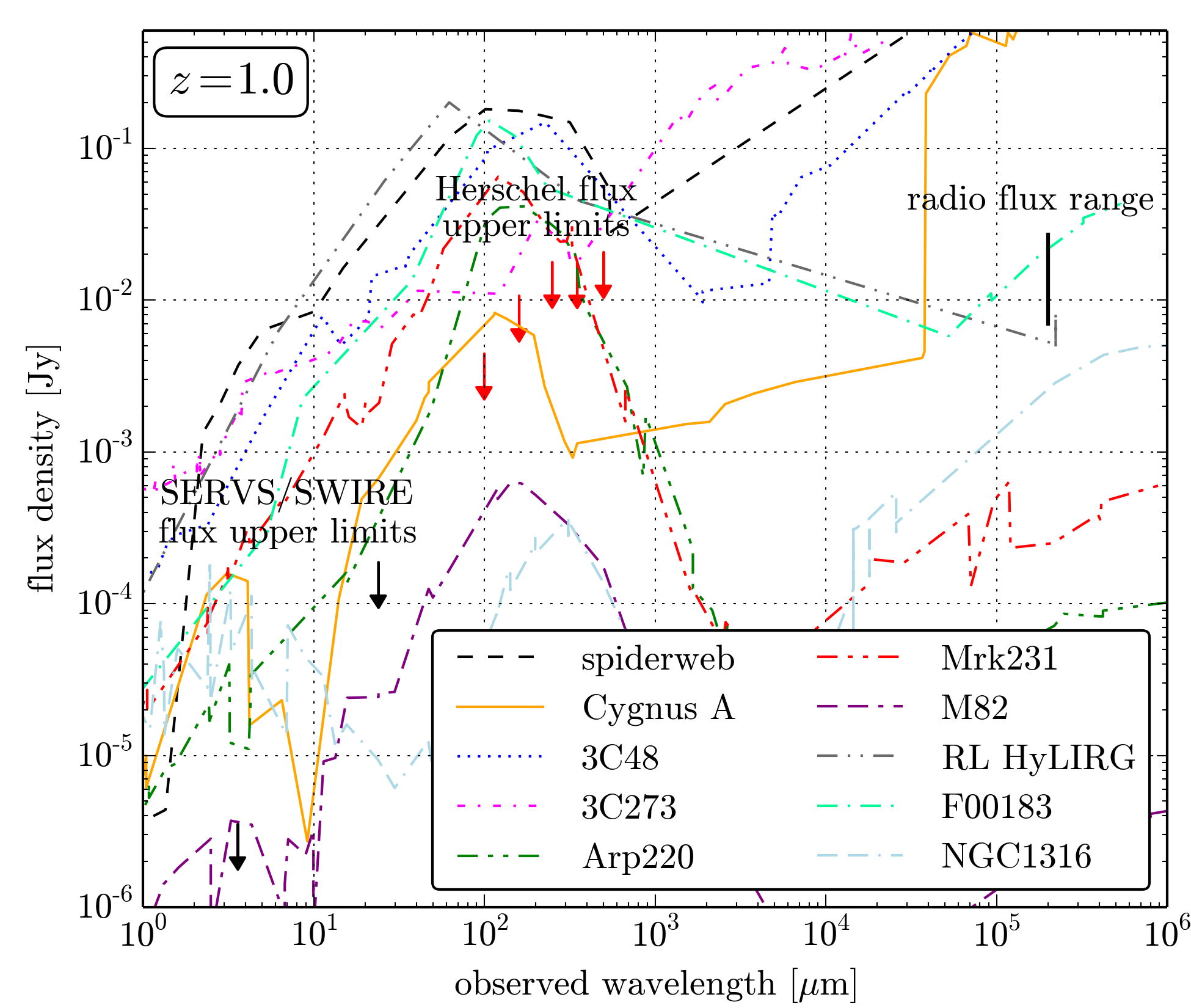}
		\includegraphics[width=8.5cm]{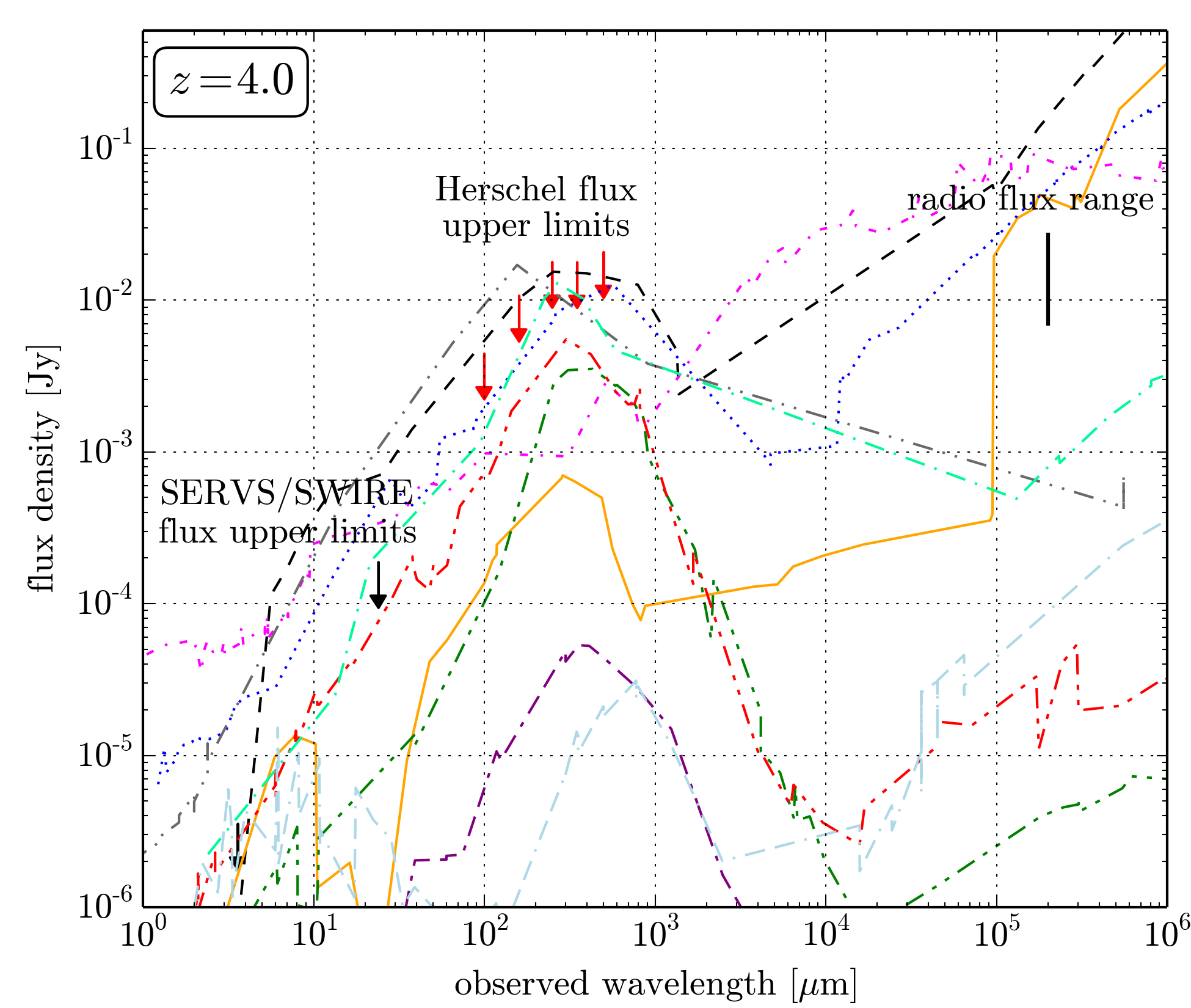}
		\includegraphics[width=8.5cm]{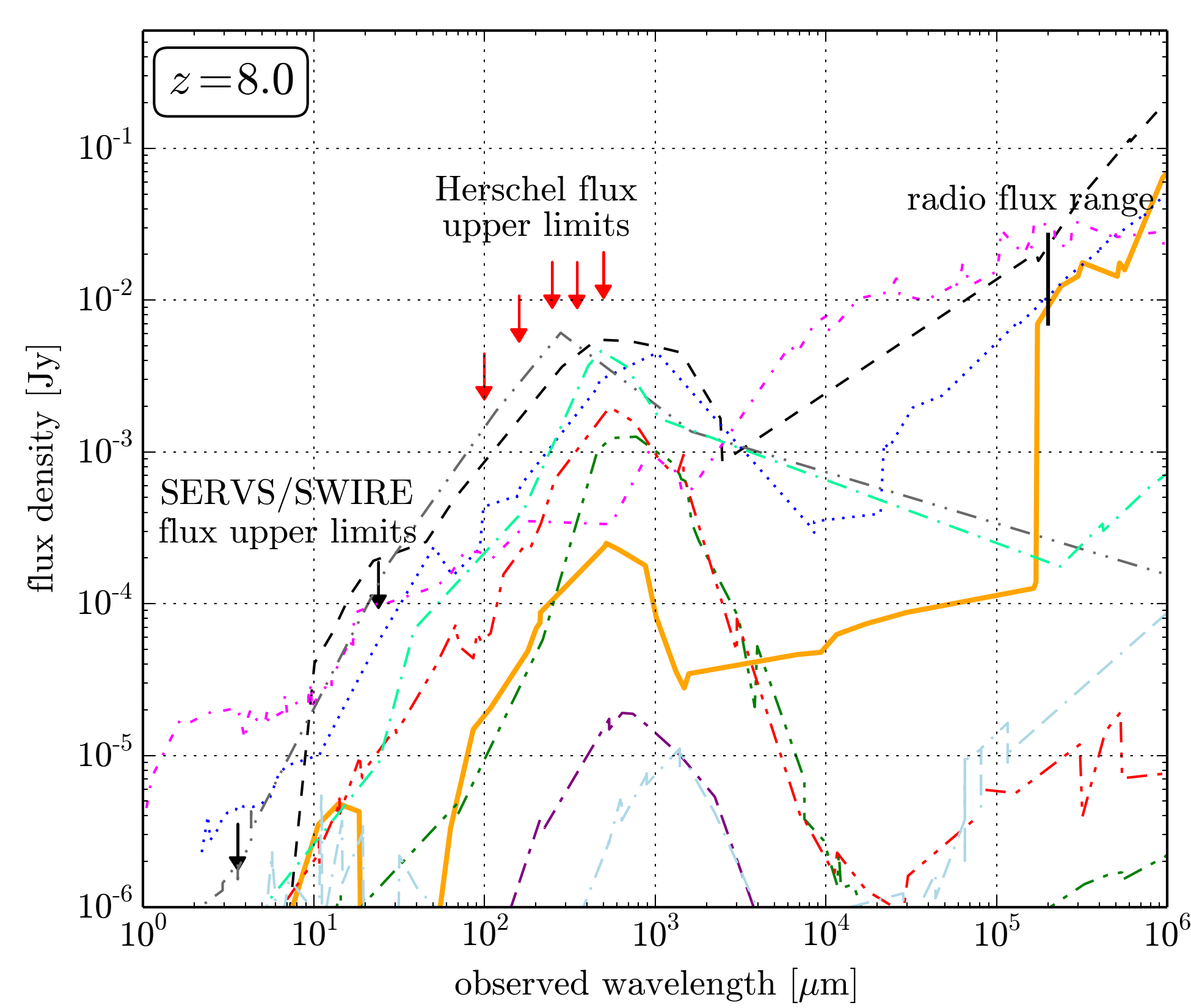}
		\includegraphics[width=8.5cm]{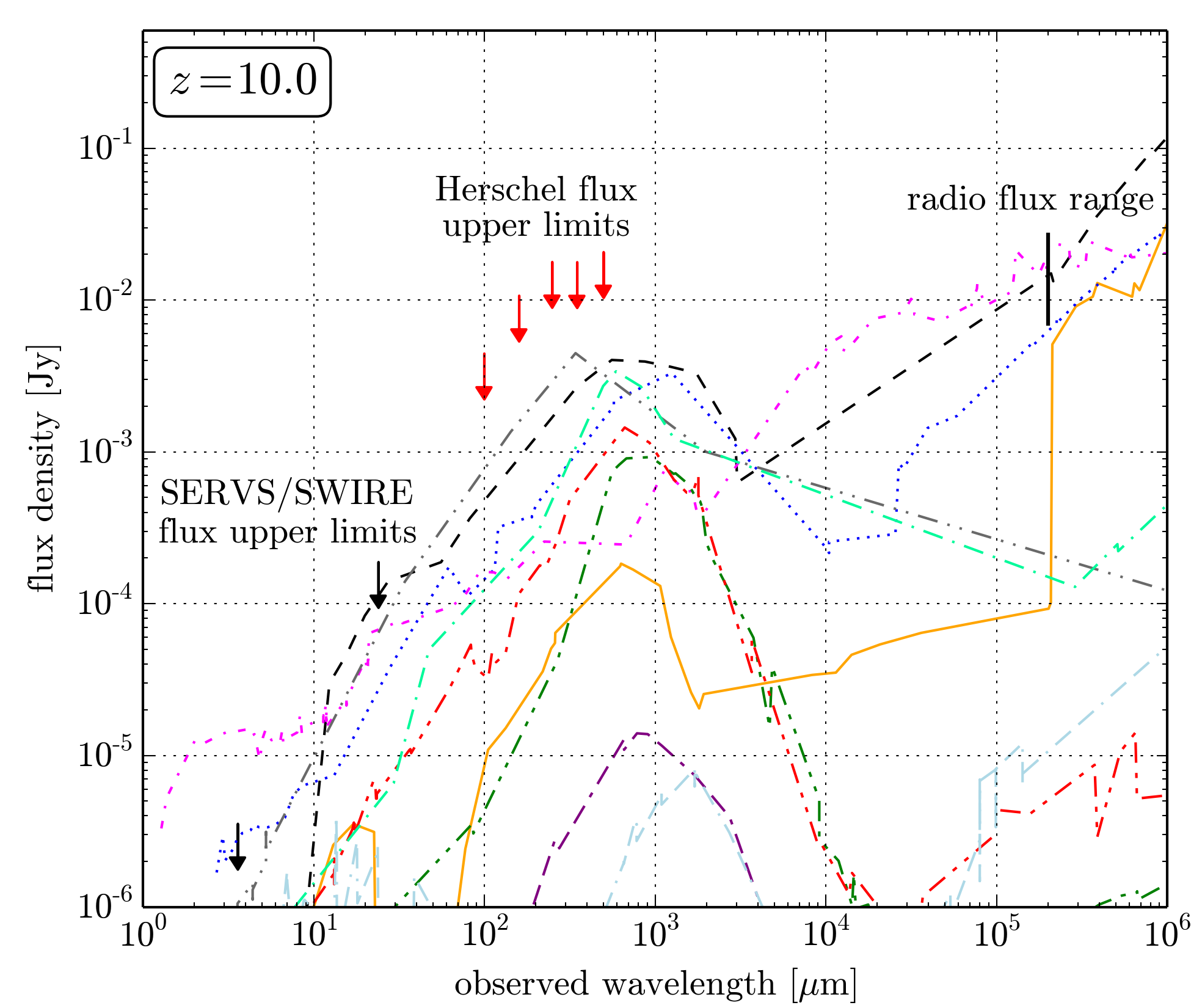}
		\caption{Broad-band SED modelling for IFRS, using the far-IR flux
		density upper limits measured in this work (red arrows), SERVS/SWIRE near- and
		mid-IR flux density upper limits (black arrows), and the detections at 1.4\,GHz
		(black bar). The template SEDs were shifted to redshifts between $z=1$ and $z=12$.
		Shown are the results at $z=1$ (upper left subplot), $z=4$ (upper right),
		$z=8$ (lower left), and $z=10$ (lower right), respectively. The legend shown in the
		upper left subplot is valid for all four subplots. At redshifts $z \lesssim
		5$, no template is in agreement with the photometric constraints of IFRS. However, if shifted to the redshift regime
		$5 \lesssim z \lesssim 8.5$, Cygnus~A (orange solid line) is in agreement with
		all constraints. The spiderweb galaxy (black dashed line) provides an appropriate
		template if shifted to very high redshifts $z\gtrsim 10.5$. The CSS source
		3C\,48 (blue dotted line) and the RL quasar 3C\,273 (magenta
		dashed-dotted line) fulfil the constraint in the radio regime at redshifts
		$6\lesssim z \lesssim 9$ and $8\lesssim z \lesssim 12$, respectively, but
		disagree with the 3.6\,$\mu$m flux density upper limits at these redshifts.}
	\label{fig:SEDmodelling_shifted}
\end{figure*}
The resulting plots for redshifts of 1, 4, 8, and 10, respectively, are
shown in Fig.~\ref{fig:SEDmodelling_shifted}.\par

In the redshift regime $1\lesssim z \lesssim 5$, no template fulfils all
available photometric constraints, i.e.\ none of these objects would produce the
observational characteristics of IFRS if placed at redshifts $z \lesssim 5$.
However, for redshifts in the range $5 \lesssim z \lesssim 8.5$, Cygnus~A
provided an appropriate template to fulfil the photometric constraints of IFRS.
The spiderweb galaxy SED was in agreement with all available photometric data of
IFRS when shifted to $z\gtrsim 10.5$. At lower redshifts, the radio and the
near-, mid-, and far-IR flux densities of the spiderweb galaxy exceeded the
measured fluxes of IFRS. We also tried templates of other HzRGs (e.g.\ 3C\,470,
4C\,23.56) and found the same qualitative result.\par

The CSS~source 3C\,48 and the RL~quasar 3C\,273 matched the measured
radio flux density of IFRS if shifted to the redshift regimes $6\lesssim z
\lesssim 9$ and $8\lesssim z \lesssim 12$, respectively. However, at these
redshifts, their 3.6\,$\mu$m flux densities exceeded the related
measured flux density upper limits of IFRS up to a factor of ten, ruling out
these templates. All other templates were found to be in strong
disagreement with the IFRS data at all redshifts between 1 and 12. Either the
radio flux densities of the templates exceeded the measured flux density of IFRS
or the templates disagreed with the near-, mid-, and far-IR flux density upper
limits of IFRS by several orders of magnitude.\par

We also estimated a score for each SED template at each redshift. For this
purpose, we derived the score~$s$ based on a modified $L_1$ norm~(e.g.\
\citealp{Horn1985}) given by
\begin{equation}
s = 10^{-\sum \limits_{i} \big| \log_{10} \frac{y_i}{m_i} \big| ~-~ \sum
\limits_{j} H[m_j - 3\sigma_j] ~ \big| \log_{10} \frac{3\sigma_j}{m_j} \big| }~.
\label{eq:score}
\end{equation}
The first sum in the exponent runs over the wavelengths~$i$ at which our sample
of IFRS is detected, where $y$ denotes the flux density of IFRS and $m$ the flux
density of the respective SED model. The second sum accounts for deviations of
the SED model from the flux density upper limits of IFRS and sums over the
wavelengths~$j$ at which the IFRS are undetected with an rms~$\sigma$, using a
Heaviside step function~$H$. In order to avoid an overweighting of the FIR
regime when deriving a score due to the five independent flux density
measurements in this regime, we only used one out of the five \textit{Herschel}
wavelengths. Thus, the wavelength resulting in the highest addend $|\log_{10}
(3\sigma_j / m_j)|$ was used for this purpose.\par

This implies that a score of 1 is obtained by an SED which is in agreement with
all eight photometric constraints of the IFRS considered in this work. We
emphasise that a score of 1 does not imply that the SED of IFRS necessarily
follows the respective template. However, the score provides useful insights how
well a template agrees with the data of IFRS at a respective redshift.
\begin{figure*}
	\centering
		\includegraphics[width=8.5cm]{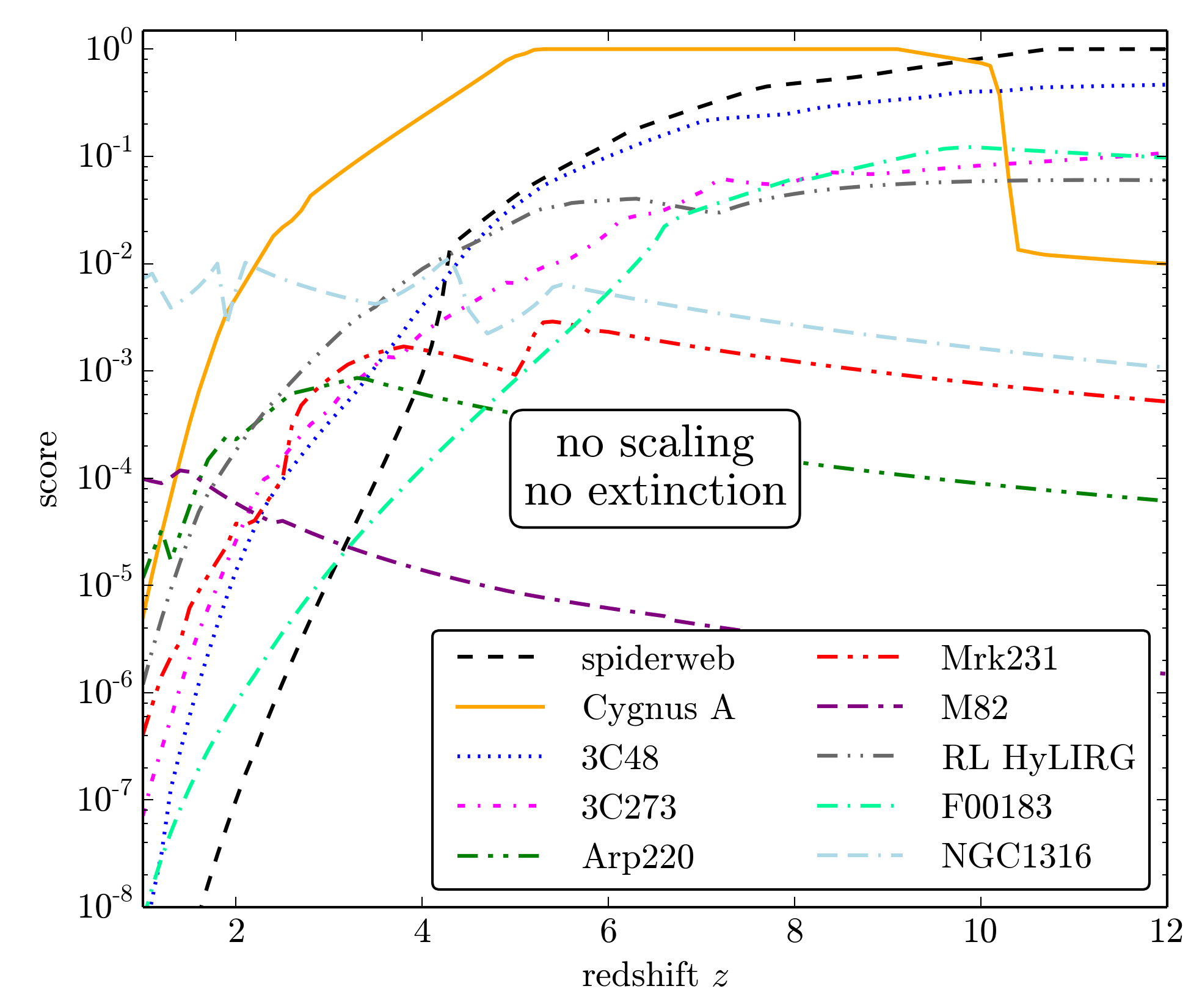}
		\includegraphics[width=8.5cm]{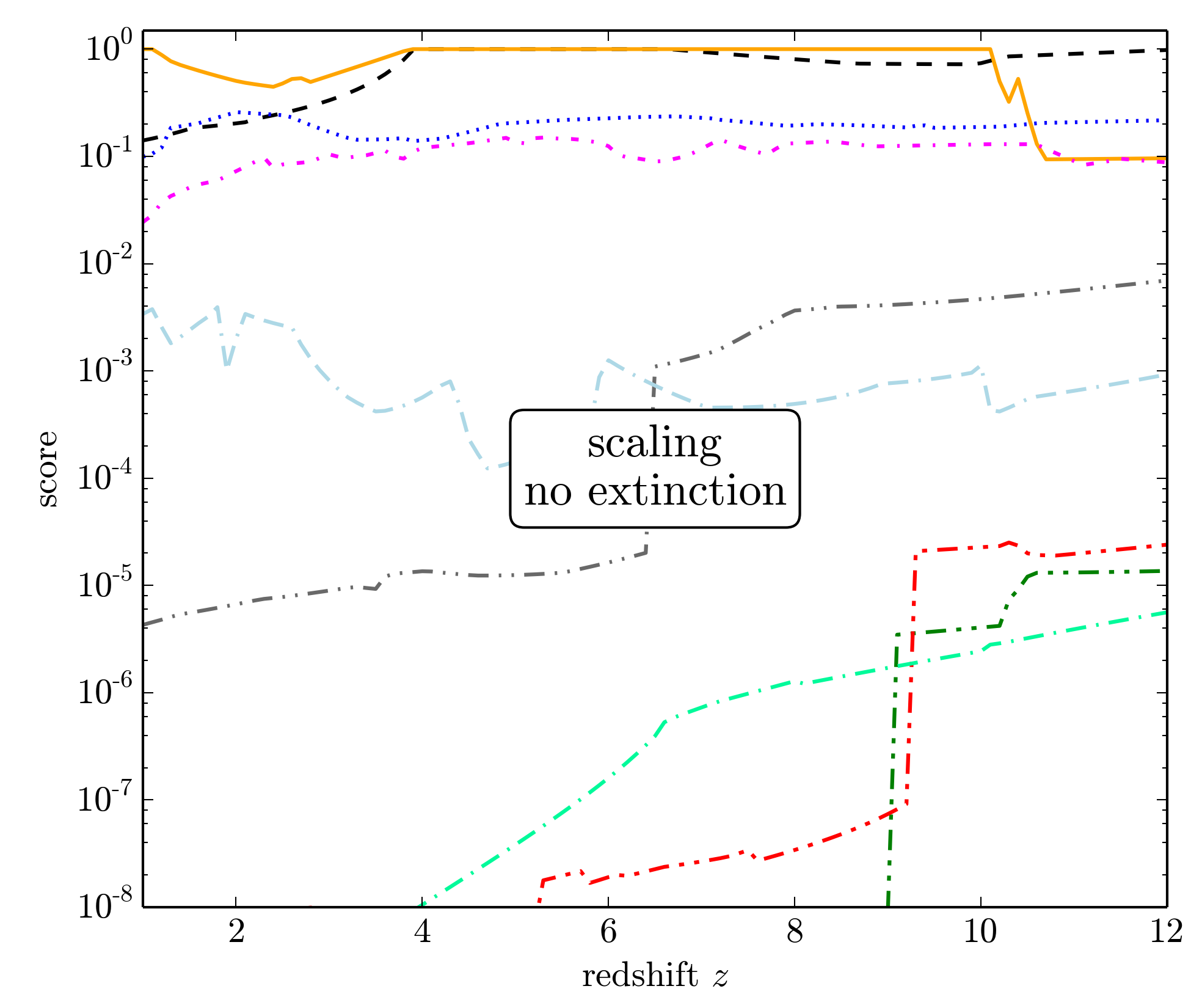}
		\includegraphics[width=8.5cm]{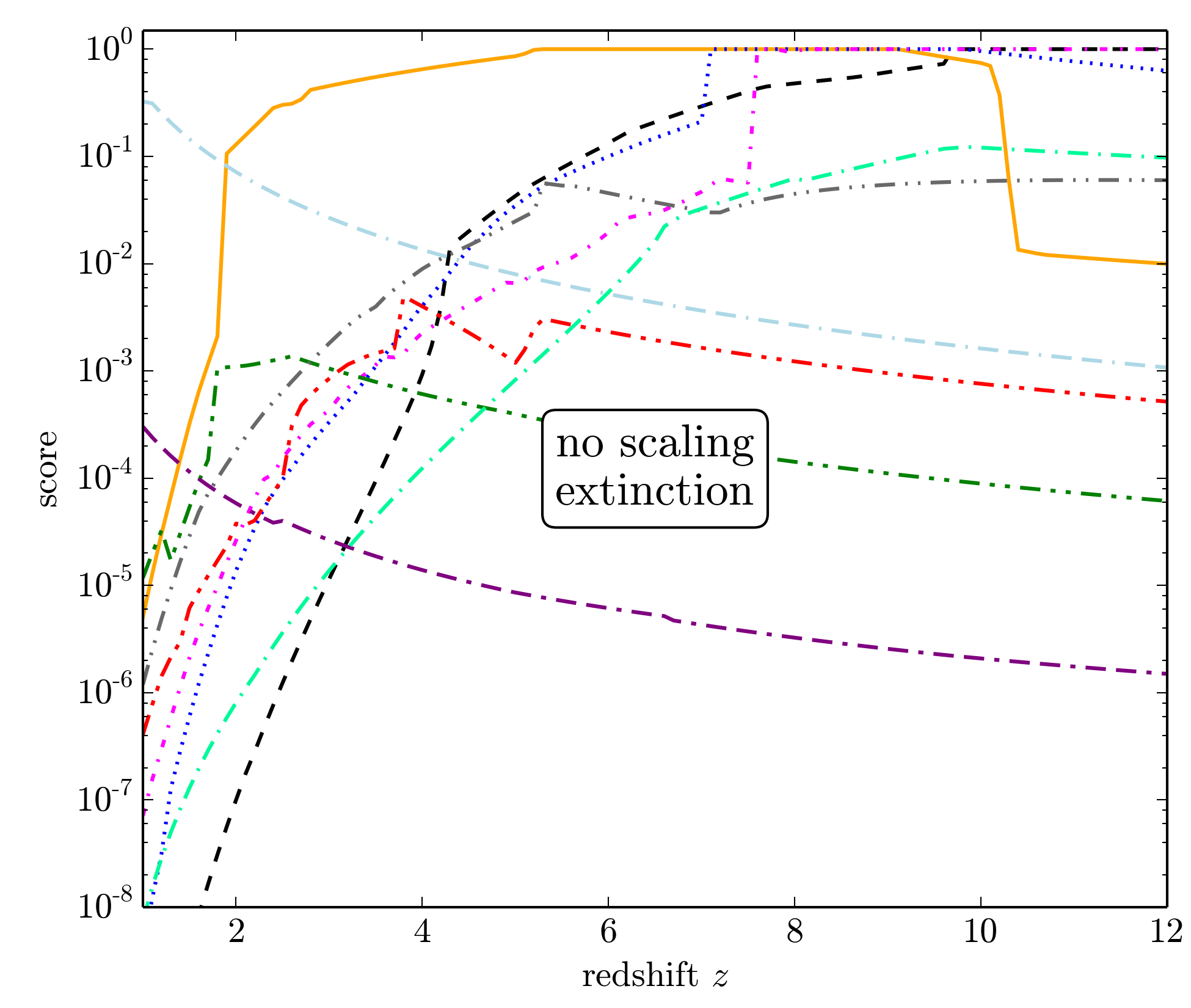}
		\includegraphics[width=8.5cm]{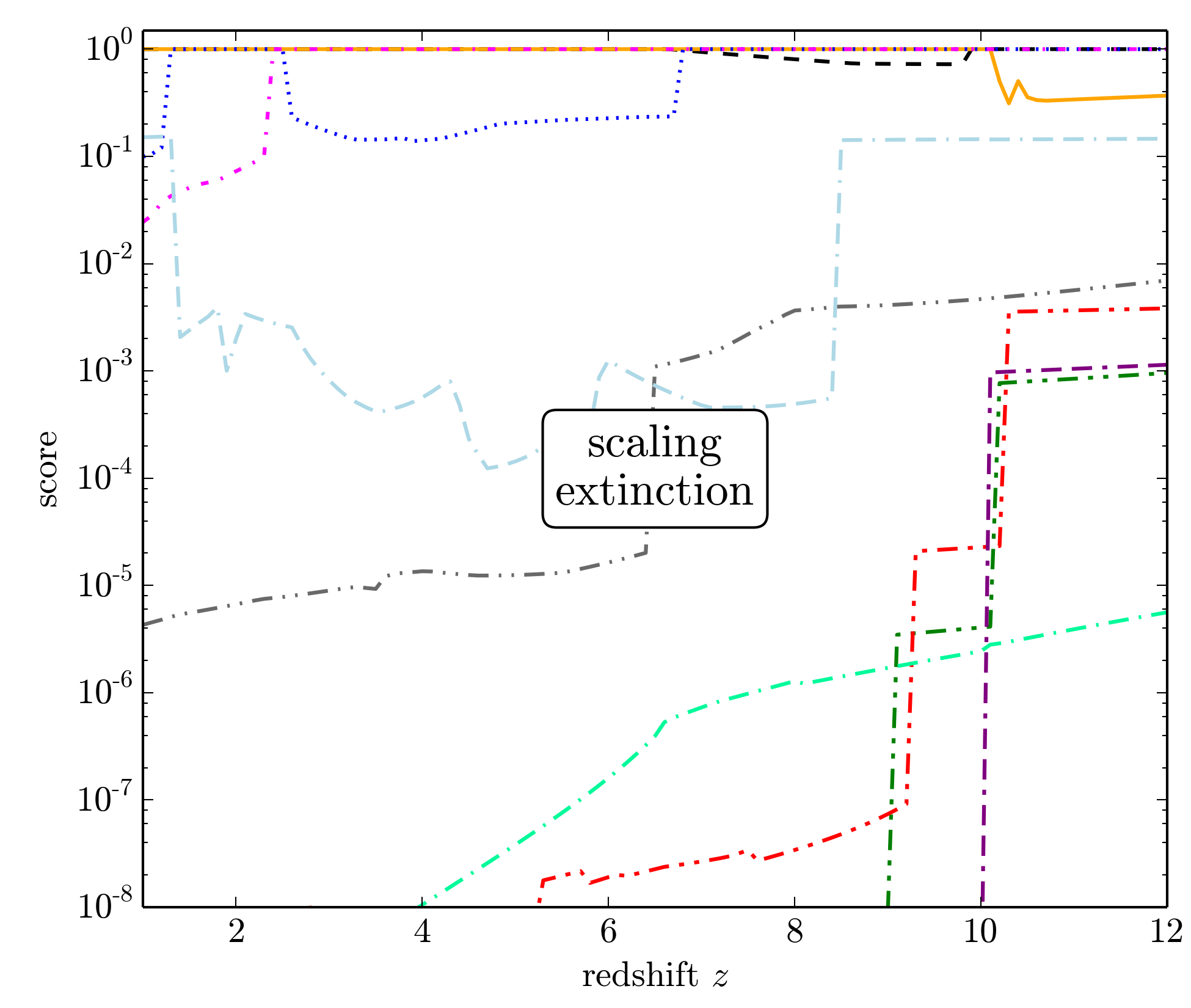}
		\caption{Score for different SED templates as a
		function of redshift, measured in our broad-band SED modelling. A
		score of 1 implies that the SED is in agreement with all photometric
		constraints of the IFRS discussed in this work. Shown is the score
		resulting from modelling without scaling and without additional extinction
		(upper left), with scaling and without additional extinction (upper right),
		without scaling and with additional extinction (lower left), and with
		scaling and with additional extinction (lower right subplot), respectively. In
		both plots on the right hand side, the templates were scaled to a 1.4\,GHz flux
		density of 15\,mJy. The legend shown in the upper left subplot is valid for
		all four subplots.}
	\label{fig:probability}
\end{figure*}
The upper left subplot in Fig.~\ref{fig:probability} shows the score as
a function of redshift for different templates.\par

We found only the spiderweb galaxy and Cygnus~A are consistent with the
photometric data of IFRS when shifted to appropriate redshifts. If IFRS
have the SED of a known galaxy, IFRS have to be at redshifts
$z\gtrsim 5$. This is consistent with the suggestion by \citet{Norris2011},
\citet{Collier2014}, and \citet{Herzog2014} that the IFRS observed with
\textit{Herschel} are at redshifts $z\gtrsim 4$ because of their faintness at
3.6\,$\mu$m as discussed above.\par

\subsubsection{Modifying broad-band SED templates in redshift and luminosity}
\label{scaling}

In the broad-band SED modelling presented in
Sect.~\ref{differentredshift}, redshift was the only free parameter. We found
that only HzRGs (e.g.\ the spiderweb galaxy) and Cygnus~A provide appropriate
templates to explain the emission characteristics of IFRS, and only if shifted
to redshifts $z\gtrsim 5$. In the following, however, we aimed at testing
whether the SED characteristics of IFRS can be reproduced by objects at lower
redshifts. For this purpose, we broadened the parameter space and used templates
modified by varying their luminosity.\par

In this approach, we included a wavelength-independent luminosity
scaling factor as an additional parameter. This allowed the
templates to be scaled to match the mean radio flux density of the IFRS
observed with \textit{Herschel} ($S_{1.4\,\mathrm{GHz}} \approx 15\,$mJy), since
these sources are solely detected in the radio regime apart from the weak
3.6\,$\mu$m counterpart of S427.\par

The upper right subplot in Fig.~\ref{fig:probability} shows the score versus
redshift for the different templates when they were scaled in luminosity to
match the median radio flux density of the IFRS discussed in this work. In this
approach, we found the spiderweb galaxy is in agreement with the available
photometric constraints at redshifts in the range $4 \lesssim z \lesssim 6.5$
when scaled in luminosity by factors between 0.08 and 0.4. At higher redshifts,
the scaled template SED exceeded the 24\,$\mu$m flux density upper limits and
therefore resulted in a score below 1. The SED template of Cygnus~A was also
found to be in agreement with the data at $z\sim 1$ and in the redshift
range $4\lesssim z \lesssim 10$ when scaled in luminosity by factors between
0.007 and 2. In the redshift range $1.5\lesssim z \lesssim 4$, the scaled
template disagreed with the 3.6\,$\mu$m flux density upper limit. All other
templates were in disagreement with at least one upper limit at any
redshift.\par

When scaling the templates to match the faint end of the observed radio
flux density range of IFRS of 7\,mJy, the spiderweb galaxy template was in
agreement with all constraints when shifted to the redshift range $ z \gtrsim
3.5$, and Cygnus~A if shifted to the range $z\lesssim 2$ or
$3\lesssim z \lesssim 10.5$. None of the other templates agreed with all photometric constraints. In contrast, if
the templates were scaled to the radio flux density of the radio-brightest IFRS
in our sample (26\,mJy), only the template of Cygnus~A was in agreement with the IR
flux density upper limits ($4.5\lesssim z \lesssim 10$). However, at $z\sim 4$,
the template of the spiderweb galaxy only slighly violated the 24\,$\mu$m flux
density upper limit.\par

\subsubsection{Modifying broad-band SED templates in redshift and extinction}
\label{extinction}

We found in Sect.~\ref{differentredshift} that the redshifted SED
templates of 3C\,48, and 3C\,273 were in agreement with the radio flux
density range and the FIR flux density upper limits of IFRS, however
exceeding the near- or mid-IR flux density upper limits. Therefore, here,
we tested whether some of our templates could be squeezed below these IR flux
density upper limits when additional extinction was added to the templates.\par

For this purpose, we added extinction in the rest-frame optical and
near-IR regime of the template SEDs if required. We used the
\citet{Calzetti2000} reddening law which reduces emission at rest-frame
wavelengths between 0.12\,$\mu$m and 2.2\,$\mu$m. We limited the additional
extinction to an arbitrary number of 500\,mag. In our modelling, we applied
energy conservation by calculating the power hidden by the additional amount of
dust in the rest-frame optical and near-IR regime and re-radiating this power at
mid- and far-IR wavelengths. We implemented this power conservation in our
modelling by adding a dust emission component of the same power as hidden by the
additional extinction in the optical and near-IR regime.\par

The emission of dust is usually described by a modified black-body spectrum,
i.e.\ $S_\nu \propto B_\nu (T) \epsilon_\nu$, where $B_\nu (T)$ is the
black-body emission given by the temperature~$T$ and $\epsilon_\nu$ is the
emissivity function at a frequency~$\nu$. We followed simple approaches, using only
one dust temperature $T = 70$\,K and assuming $\epsilon \propto \nu^\beta$ with
$\beta = 1.5$. The emissivity spectral index~$\beta$ depends on dust grain
properties as size and composition and usually ranges between 1 and 2. We note
that the choice of this spectral index in the given range did not
qualitatively change our results; neither did changing the dust peak temperature
between 30\,K and 100\,K.\par

The plot of the scores resulting from this modelling approach---adding
extinction to the redshifted SED templates---is shown in the lower left subplot
in Fig.~\ref{fig:probability}. It should be noted that no scaling in luminosity
as presented in Sect.~\ref{scaling} had been applied in this approach. We found
the spiderweb galaxy template in agreement with the data at redshifts $z\gtrsim
9.5$. At $z\sim 10$, 5.6\,mag of additional extinction brought this SED template
in agreement with the data. At higher redshifts, no additional extinction was
required. Cygnus~A was in agreement with all flux density upper limits
in the redshift range $5 \lesssim z \lesssim 9$ without adding extinction as
shown in Sect.~\ref{differentredshift}. 3C\,273 matched the data at redshifts
$z\gtrsim 7.5$ if between 8.6\,mag and 1.4\,mag of extinction were added. 3C\,48
was in agreement with all IR flux density upper limits at $z\gtrsim 7$ if
between 1.4\,mag and 0.2\,mag of extinction were added. However, the radio flux
density of 3C\,48 deceeded the minimum flux density of 7\,mJy at high redshifts
$z\gtrsim 9.5$. All other SED templates were found to be in disagreement with
the data at any redshift.\par

Even enormous and unphysically high amounts of extinction of 500\,mag
were insufficient to squeeze the near-IR emission of some templates below the
flux density upper limit at the observed wavelength of 3.6\,$\mu$m. In that
case, we did not add extinction to the templates, resulting in the appropriately low score at the
respective redshift. This happened for all radio-quiet templates like Arp220,
Mrk231, or M82 at redshifts~$z \lesssim 5$. However, at higher redshifts, when
additional extinction was able to squeeze the templates below the IR flux
density upper limits, the templates were by far too faint in the radio regime to
explain the emission features of IFRS.\par

\subsubsection{Modifying broad-band SED templates in redshift, luminosity, and
extinction}
\label{scalingextinction}

Finally, in the last step of our broad-band SED modelling, we combined
the approaches presented in Sections~\ref{scaling} and \ref{extinction}. Here,
we modified the SED templates by shifting them in the redshift range $1\leq
z\leq 12$, scaling them in luminosity to match the observed radio flux density,
and adding extinction---if required---to squeeze the SED templates below the
near- and mid-IR flux density upper limits. IR re-radiation was implemented
as described in Sect.~\ref{extinction}.\par

The resulting plot of the score as a function of redshift
is shown in the lower right subplot in Fig.~\ref{fig:probability}. We note that
each template at each redshift had been modified by an individual scaling factor
and---if necessary---an individual amount of extinction had been added.\par

Similar to the approaches presented before, we only found the spiderweb
galaxy, Cygnus~A, the CSS source~3C\,48, and the RL quasar~3C\,273 to provide
suitable templates to explain the observational data of IFRS. The spiderweb
galaxy was in agreement with the data at all redshifts except for the range
$6.5\lesssim z \lesssim 9.5$. In this redshift range, the template exceeded the
24\,$\mu$m flux density upper limit despite adding additional extinction.
The template of Cygnus~A matched all constraints at $z\lesssim 10$ and
3C\,273 at redshifts $z\gtrsim 2.5$. 3C\,48 was in agreement with the data at any redshift
except for very low redshifts $z\lesssim 1.5$ and except for the range
$2.5\lesssim z \lesssim 7$.\par

In the respective redshift ranges, an appropriate scaling factor and an
appropriate amount of extinction was found for the spiderweb galaxy, Cygnus~A,
3C\,48, and 3C\,273 to match all photometric constraints known for these IFRS.
For example at $z=4$, the spiderweb galaxy and Cygnus~A needed to be scaled down
in luminosity by a factor of 10 and 3, respectively, to match the
radio flux density of 15\,mJy. No extinction was needed in these cases. In
contrast, 3C\,273 required 3.2\,mag of extinction when scaled down by a factor
of 5 to match the radio data point, while 3C\,48 could not be modified at this
redshift to match all constraints.\par

Matching the radio flux densities of the radio-brightest IFRS in our sample
(26\,mJy at 1.4\,GHz), again in case of a shift to $z=4$, an additional
extinction of 1.3\,mag was required in case of Cygnus~A, whereas no
reasonable amount of extinction could be added to the templates of the spiderweb
galaxy, 3C\,48, and 3C\,273, to match the near- and mid-IR flux density upper
limits. Considering the radio-brightest IFRS in our sample, at $z\lesssim 4.5$,
only the Cygnus~A SED template could be brought in agreement with the
constraints of these most extreme---i.e.\ with the highest radio-to-IR flux
density ratios---IFRS, if up to 4.1\,mag of extinction were added.
At higher redshifts $z\gtrsim 10$, 3C\,48 and the spiderweb galaxy
provided appropriate templates, too. 3C\,273 was in agreement with the data at
$4.5\lesssim z \lesssim 5.5$ and $7.5\lesssim z \leq 12$. In contrast, if the
templates were scaled to a 1.4\,GHz flux density of 7\,mJy, the spiderweb
galaxy, Cygnus~A, 3C\,273, and 3C\,48 could be modified to be in agreement with
all photometric constraints at any redshift, except for Cygnus~A
at very high redshifts $z\gtrsim 10$.\par

All other templates used in this broad-band SED modelling disagreed with the
observational data when shifted to the redshift range $1 \leq z \leq 12$, scaled
in luminosity, and extinction was added. These templates were clearly ruled out
by their disagreement with the flux density upper limits of IFRS in the FIR
regime measured in this work. Particularly, the templates of an RL
HyLIRG or F00183, representing common classes of luminous objects at $z\gtrsim 2$, were
found to be inappropriate to reproduce the SED constraints of IFRS at any
redshift.\par

\subsubsection{Summary of broad-band SED modelling}
\label{summary_broadband}

We summarise our findings from the broad-band SED modelling as follows.
When using templates of existing galaxies (Sect.~\ref{differentredshift}), the
SED characteristics of IFRS can only be explained by the SEDs of Cygnus~A and
HzRGs, however only at high redshifts $5\lesssim z\lesssim 8.5$ and $z\gtrsim
10.5$, respectively. If IFRS are at lower redshifts ($z\lesssim 5$), their SED
constraints can be fulfilled by fainter versions of an HzRG or Cygnus~A (Sects.
\ref{scaling} and \ref{scalingextinction}), by a dust-obscured CSS source, or by
a dust-obscured RL quasar (Sects.~\ref{extinction} and
\ref{scalingextinction}).\par

If IFRS are linked to HzRGs as suggested by \citet{Middelberg2011},
\citet{Norris2011}, \citet{Collier2014}, and \citet{Herzog2014}, there are two
different suggested options: (a) IFRS are very similar to HzRGs---i.e. with
similar luminosities---however at higher redshifts ($z\gtrsim 5$), or (b) IFRS
are fainter siblings of HzRGs at similar redshifts ($1 \leq z \lesssim 5$).
Based on the template of the spiderweb galaxy, we found that both options are
consistent with the FIR flux density upper limits of IFRS measured in this work
and the related broad-band SED modelling. If IFRS have the same intrinsic
properties as HzRGs, represented by no scaling and no adding of extinction in
our modelling, they have to be located at very high redshifts $z\gtrsim 10.5$
(Fig.~\ref{fig:probability}, upper left subplot). Alternatively, if IFRS are
fainter siblings of HzRGs, i.e. they are scaled down in luminosity and
potentially more dusty, they could be located at lower redshifts $z\geq
1$~(Fig.~\ref{fig:probability}, lower right subplot). It should be noted that
extinctions between 14\,mag and 5\,mag had to be added to the scaled
spiderweb galaxy SED template at redshifts $z\leq 2$ to match the IR flux density upper
limits.\par

Two of the templates, 3C\,48 and 3C\,273, were found to be in agreement with the
data of IFRS only if additional extinction in the order of up to several
magnitudes was added to the templates as discussed in
Sect.~\ref{scalingextinction}. Particularly, these modified templates did not
exceed the FIR flux density upper limits of IFRS measured in this work. This
implies that a significant dust obscuration in IFRS, explaining the optical and
near-IR faintness of these objects, cannot be ruled out. In fact, if the IFRS
observed here are similar to 3C\,48 or 3C\,273, these sources have to be dust
obscured to produce the photometric constraints of IFRS.\par

The results of this broad-band SED modelling are generally in agreement with the
findings by \citet{Herzog2014}. \citeauthor{Herzog2014} measured spectroscopic
redshifts of three IFRS detected in the optical and near- and mid-IR regime and
did the first redshift-based SED modelling for the class of IFRS. They find the
modified templates of 3C\,48 and 3C\,273 in agreement with all available
photometric data of these three IFRS, whereas the templates of star forming
galaxies and Seyfert galaxies were clearly ruled out. We also agree with the
finding that only RL objects can explain the photometric constraints of
IFRS. However, \citeauthor{Herzog2014} find that no
additional extinction is required for 3C\,48 and 3C\,273 in their SED
modelling. Here, in contrast, we found that IFRS might be dust obscured RL
quasars with spectra similar to 3C\,48 or 3C\,273. However,
\citeauthor{Herzog2014} studied the IR-brightest IFRS, whereas the IR-faintest
IFRS are studied in the current work, providing a valid reason for the
discrepancy. \citet{Huynh2010} modelled the SED of individual IFRS and find
that the scaled SED template of 3C\,273 is in agreement with their data if
around one magnitude of extinction was added. This in agreement with
our modelling, considering that the IFRS from \citeauthor{Huynh2010}
are more than ten times fainter in the radio compared to the IFRS discussed in the present work.
We found that IFRS have to be dust obscured, too, if IFRS are similar to
HzRGs at $z\lesssim 3.5$. However, if the spectral shape of IFRS is related to
that of HzRGs, no extinction is required to explain the SED characteristics of
IFRS if these templates are shifted to higher redshifts.\par

Our broad-band SED modelling showed that the SED templates of the
spiderweb galaxy and of Cygnus~A can be shifted to $1 \leq z\leq 2$ and modified
to agree with the photometric observations of IFRS. However, in contrast, all
21~spectroscopic redshifts of IFRS are in the range $1.8 < z < 3$, i.e.\ no IFRS
has been found at lower redshifts $z\sim 1$. There are two potential reasons for
this discrepancy. (a) Our modelling---by shifting the templates in redshift,
scaling them in luminosity, and adding extinction---is not physically realistic
for galaxies at $z\sim 1$. At these low redshifts, all appropriate templates
needed to be scaled down in luminosity up to factors of 500. Presumably, the
black hole mass needs to be scaled down by a similar factor. However, RL~AGN
with lower-mass black holes are very rare (Rees et al., submitted). If the
galaxy loses its radio excess by scaling down the black hole mass, the
characteristically high radio-to-IR flux density ratio of IFRS would not be
reached and the galaxy would not be considered as an IFRS. Furthermore, in case
of the spiderweb galaxy template, the required high amounts of additional
extinction of around 10\,mag at low redshifts are very rare and decrease the
possibility to observe such extreme objects. (b) The IFRS discussed in this work
are intrinsically different to the IFRS with known spectroscopic redshifts.
Cygnus~A provided the only template which can reproduce the characteristics of
IFRS at $z\sim 1$ without additional dust obscuration. However, it is known that
Cygnus~A contains a hidden quasar~\citep{Antonucci1994}, resulting in narrow
emission lines, whereas IFRS were found to show broad emission lines in their
optical spectra~\citep{Collier2014,Herzog2014}. If the IFRS discussed in this
work are indeed fainter versions of Cygnus~A at redshifts $z\sim 1$, these IFRS
would have different properties than the IR-detected IFRS presented by
\citeauthor{Collier2014} and \citeauthor{Herzog2014}. They would form a separate
subclass with different redshift and emission line properties. However, so far,
no evidence has been found that the population of IFRS might be devided into two
sub-classes.\par

We note that the modified templates of Cygnus~A, 3C\,48, and
3C\,273---found to match all photometric constraints of IFRS---can be considered
as HzRGs in our modelling at $z\geq 1$. At these redshifts, they fulfil both
selection criteria of HzRGs: $z>1$ and $L_{\mathrm{3\,GHz}} >
10^{26}\,\mathrm{W\,Hz^{-1}}$. Although we scaled down the templates in
luminosity at lower redshifts, they were radio-luminous enough to be considered
as HzRGs. We can therefore conclude that only HzRGs can explain the
photometric characteristics of IFRS.\par

Our broad-band SED modelling showed several templates that contain a
significant SB contribution in disagreement with the available
photometric data (Arp220, M82, RL~HyLIRG, F00183). In contrast, the highly
star forming spiderweb galaxy, fulfilled the photometric data constraints of
IFRS. Therefore, in the following, we analysed a potential star forming
contribution to the SED of IFRS, instead or in addition to an RL~AGN.\par

\subsection{IR SED modelling}
\label{IRSEDmodelling}
One of the crucial questions in revealing the nature of IFRS consists in
unmasking the star forming activity and its contribution to the energy output in
contrast to the activity of the galactic nucleus. In the IR regime, the emission
of galaxies is mainly given by these two components, i.e.\ AGN and SB. The AGN
emission comes from dust, reprocessing far-UV through optical light, and peaks
at around 10\,$\mu$m, corresponding to a dust temperature of 300\,K. In
contrast, the stellar component is strongest in the FIR and arises from stellar
emission reprocessed to the FIR~regime by dust grains with a maximum at a
wavelength around 100\,$\mu$m, corresponding to 30\,K.\par

\subsubsection{IR SED fitting based on IR flux density upper limits}
\label{IR_SED_modelling_IRonly}

In order to decompose the IR~emission of HzRGs, \citet{Drouart2014} fit the
IR~SED of HzRGs constrained by \textit{Herschel} and ancillary IR~data. They used an
IR~SED model based on the assumption that the IR~emission comes from dust
heated by star forming activity and an AGN. They built template SEDs, adding an
empirical AGN~template and one SB template out of a set of empirical
SB templates.\par

Here, we followed this approach, aiming at setting upper limits on the
IR~SED of IFRS based on the available FIR~flux density upper limits measured in
Sect.~\ref{photometrystacking}. We used templates derived by
\citet{Mullaney2011}, covering the rest-frame wavelength range between 6\,$\mu$m
and 1090\,$\mu$m. \citeauthor{Mullaney2011} empirically built five different
SB~templates, differing in their peaking temperature as well as in the strength
of the emission of the polycyclic aromatic hydrocarbon~(PAH) molecules. These
templates cover the entire range of host galaxies~(see \citealp{Mullaney2011}
for a detailed description of the SEDs). Based on these SB~templates, they
derived one AGN~template as the residual SED after removing the
SB~contribution.\par

We used these templates and built the total IR~SED by multiplying the
AGN~template by a wavelength-independent luminosity scaling factor and adding
one SB~template which was also multiplied by a wavelength-independent luminosity
scaling factor. We set an upper limit on the IR emission by modelling the SED in
order to minimise the deviation between the total~SED, i.e.\ the sum of SB and
AGN template, and the available observed $3\sigma$ FIR~flux density upper limits
from the stacked maps (see last column in Table~\ref{tab:FluxResults}).
It was required that the available flux density upper limits were fulfilled in
the modelling. In this procedure, the most appropriate SB~template out of the
five available templates was determined based on the lowest deviation. At
redshifts $z\leq 3$, the observed 24\,$\mu$m emission fell into the templates'
wavelength coverage. At these redshifts, we used the $3\sigma$ flux
density upper limit at 24\,$\mu$m---measured in Sect.~\ref{stackingIR}
from the stacked map---in addition to the FIR~flux density upper limits measured
in the present work.\par

By this approach, we determined exactly that total IR~SED---consisting
of the AGN~template and one SB~template---which has maximum IR flux but is
still in agreement with all available flux density upper limits.
\begin{figure}
	\centering
		\includegraphics[width=\hsize]{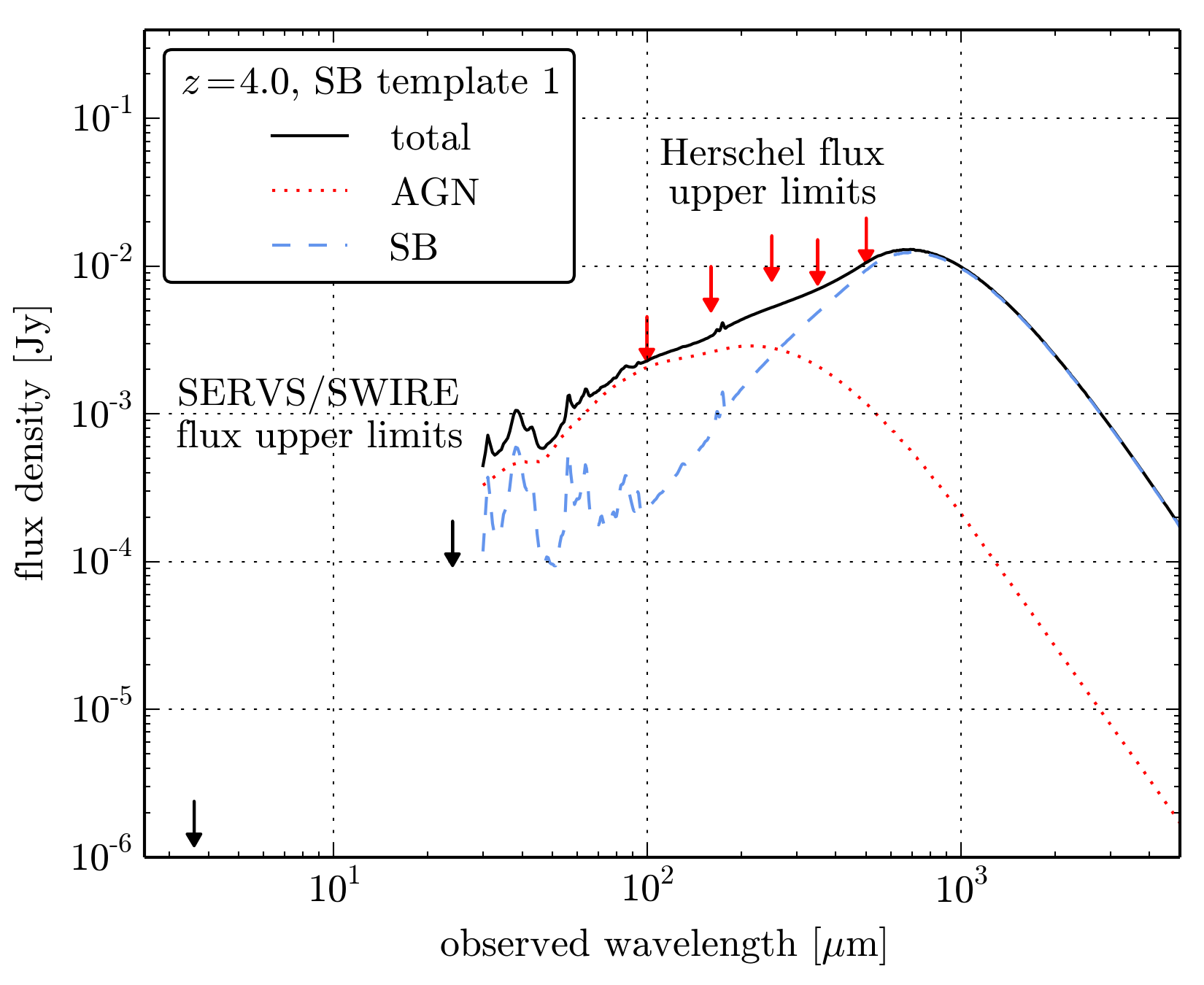}
		\caption{IR~SED~modelling for IFRS at redshift $z=4$, using the
		far-IR flux density upper limits (red arrows) measured in this work. The total~(black
		solid line) IR~emission is composed of an AGN~component~(red dotted line) and
		a SB component~(blue dashed line). The maximum model represents
		an upper limit on the IR~emission of IFRS and is defined by the
		highest IR flux which is in agreement with all available flux
		density upper limits.}
	\label{fig:IRSEDfit}
\end{figure}
Figure~\ref{fig:IRSEDfit} shows the resulting IR~SED~modelling for
$z=4$. We emphasise that our modelling represents an upper
limit on the total IR emission of IFRS.\par

This approach implies that the true AGN~contribution to the emission of the IFRS
could be higher than the contribution calculated from the maximum total
model described above, though accompanied by a lower SB~contribution,
and the other way around. However, the difference to the highest possible
contribution of each component is generally rather low.\par


By this modelling, aiming at setting upper limits on the IR emission
and referred to as ``maximum model'', we decomposed the maximum IR~SED into an
AGN~component and an SB~component.
Based on this decomposition, we were able to derive IR~luminosities of both
components by integrating the flux density in the rest-frame wavelength range
between $8\,\mu$m and $1000\,\mu$m.
\begin{figure}
	\centering
		\includegraphics[width=\hsize]{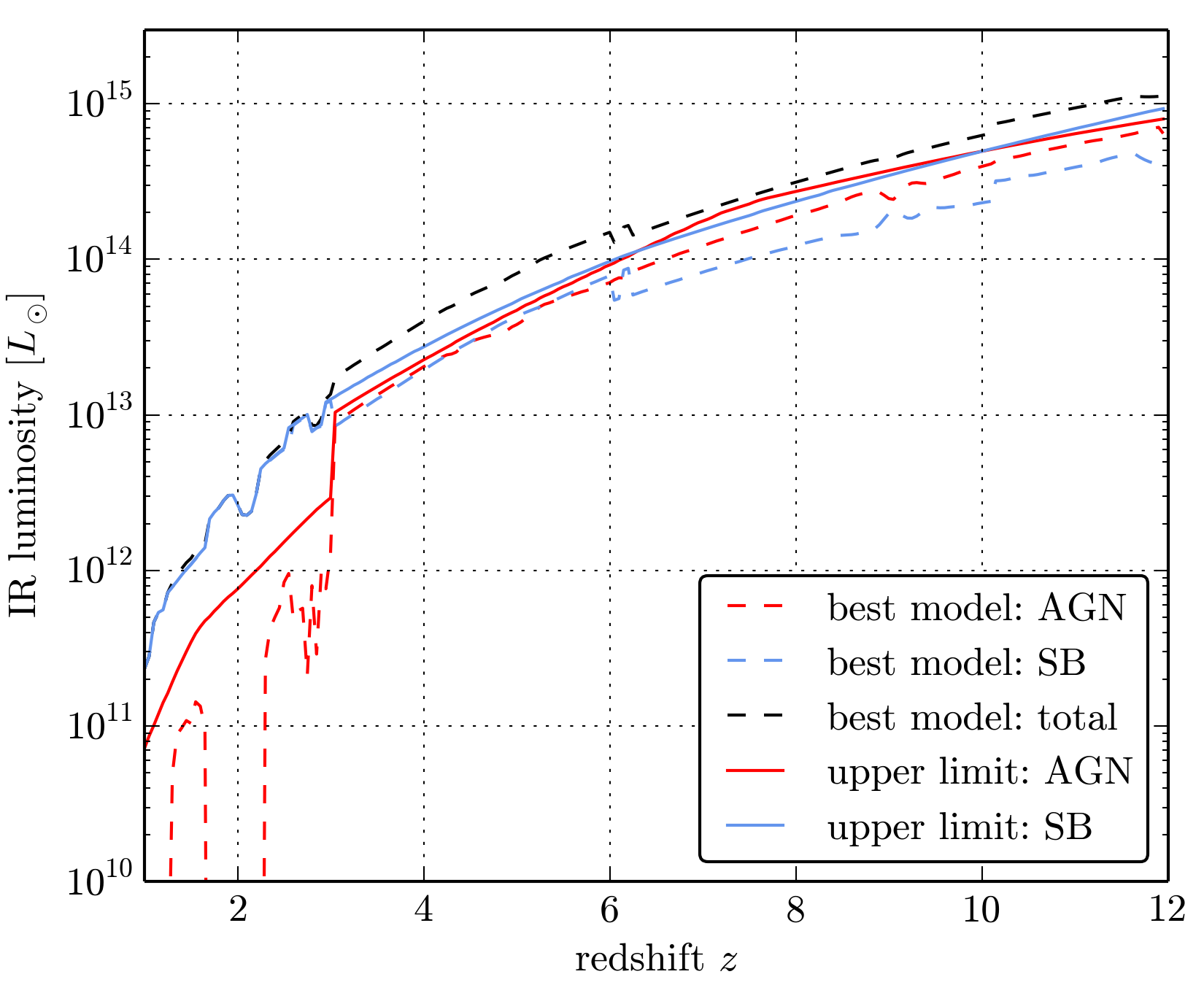}
		\caption{Infrared luminosity versus redshift for the SB~(blue) and AGN~(red)
		component as well as for the total IR luminosity (black). The dashed lines
		show the data of the maximum IR SED model, combining the AGN and the
		most appropriate SB template. The solid lines show the IR luminosity upper
		limits for both components, assuming that the other component does not
		contribute to the IR emission at each time.}
	\label{fig:IRluminosity-z}
\end{figure}
Figure~\ref{fig:IRluminosity-z} shows the IR~luminosities versus
redshift for both components and the total IR~luminosity of the maximum
model.
Furthermore, the figure shows the absolute upper limit for both the SB and the AGN
IR~luminosity, assuming that the other IR~emitting process does not contribute
at each time. For both components, the luminosities of the maximum
model are generally only slightly lower than the absolute upper limit luminosities,
except for the SB component at very high redshifts.
We note that the jump in the maximum IR~luminosity of the AGN and of the
SB~component at $z = 3$ is related to the flux density upper limit at 24\,$\mu$m
which only coincides with the template's wavelength range at redshifts $z\leq 3$.\par

We emphasise that almost any decomposition of the IR SED into SB and AGN
component is possible for IFRS since this wavelength regime is only constrained
by upper limits. Therefore, we are only able to set upper limits on the
luminosity. Particularly, we note that the maximum model shown in
Figures~\ref{fig:IRSEDfit} and \ref{fig:IRluminosity-z} is not more likely than
any other combination of SB and AGN components which is in agreement with the
flux density upper limits. However, the maximum model represents exactly that
combination of SB and AGN components which results in the highest total flux but
is still in agreement with all flux density upper limits. Therefore, this
maximum model sets the upper limit on the total IR luminosity.\par

We found that the FIR flux density upper limits measured in this work
constrain IFRS to have total IR~luminosities of $<10^{12}\,L_\odot$ at redshifts
$z\lesssim 1.5$. At $z\lesssim 2.5$, IFRS can have a maximum IR luminosity
between $10^{12}\,L_\odot$ and $10^{13}\,L_\odot$. Such sources are labelled as ULIRGs.
At redshifts $z \lesssim 6$, the IR luminosity of IFRS can be at most
$10^{14}\,L_\odot$ which allows IFRS to be HyLIRGs.\par

\citet{Drouart2014} find absolute numbers and upper limits for the total
IR~luminosity of HzRGs in the range of a few to a few tens of
$10^{12}\,L_\odot$. Our redshift-dependent IR~luminosity upper limits of IFRS
are in agreement with these numbers of HzRGs.\par

We note that the maximum model shown in Fig.~\ref{fig:IRluminosity-z} (dashed
lines) is unphysical for redshifts~$z\lesssim 2.5$. For these redshifts, the
AGN~contribution to the maximum model is very low, corresponding to a very low
AGN~activity or even an absent AGN. Instead, according to that model, the
IR~emission is mainly or even completely produced by SB~activity. In case of a
purely star forming galaxy, a direct connection between the radio and the
IR~emission was found~(i.e. \citealp{Yun2001}), known as radio-IR correlation.
Using the outcome of the maximum model that the emission of IFRS is dominated by
the SB~component for redshifts~$z\lesssim 2.5$, we could estimate the
radioluminosity from the IR~luminosity. Using our FIR luminosity upper limits,
we found that this radio luminosity is at least two orders of magnitude lower
than the radio luminosity calculated from the measured 1.4\,GHz flux density of
IFRS. This discrepancy clearly showed that the existence of an AGN in the IFRS
is essential to explain their radio flux densities. Modelling their IR~SED by
only an SB~template is inappropriate. The existence of an AGN in IFRS is also in
agreement with VLBI observations by \citet{Norris2007},
\citet{Middelberg2008IFRS_VLBI}, and \citet{Herzog2015a}.\par

\subsubsection{Including the radio data point into the IR SED modelling}
\label{IRmodelling+radio}
The IR SED modelling presented above, aiming at setting upper
limits on the total IR flux of IFRS, led to unphysical results because it
neglected the measured radio flux density of IFRS. In this section, we attempted
to address this issue by expanding our modelling to the radio
regime based on an RL AGN template.\par

In Sect.~\ref{broadbandmodelling}, we showed that all SEDs in agreement
with the photometric characteristics of IFRS were HzRGs and that the spiderweb
galaxy provided an appropriate SED template in the broad-band SED modelling. It is
known that both star forming activity and an AGN contribute to the emission of
this powerful galaxy. Therefore, in the following, we used the broad-band SED of
the spiderweb galaxy, built from NED and the additional IR data from
\citet{Seymour2012} as described in Sect.~\ref{broadbandmodelling}, as a basic
template, modified by a wavelength-independent luminosity scaling factor. Beyond
that, to account for an SB component independent of the component already
present in the spiderweb galaxy template, we added the most appropriate SB
template from the sample by \citet{Mullaney2011}. Similar to our approach in
Sect.~\ref{IR_SED_modelling_IRonly}, we modified this SB template by a
wavelength-independent luminosity scaling factor. However, here, we also used
the radio-IR correlation~\citep{Yun2001} to estimate the radio luminosity of the
star formation component, in addition to that from the RL~AGN.\par

Recapitulating, similar to the approach presented in
Sect.~\ref{IR_SED_modelling_IRonly}, we composed our total SED of two different
components. The spiderweb galaxy contributed SB and AGN emission, while the
additional SB component contributed additional SB emission. Based on the two
independent scaling factors, both the AGN contribution and the SB contribution
could be varied independently. We note that the scaling factor of the additional
SB component could also be negative, i.e.\ representing a lower star forming
activity than that in the spiderweb galaxy.\par

Similar to our approach in Sect.~\ref{IR_SED_modelling_IRonly}, we
modelled the maximum IR SED by maximising the flux of the total
template---i.e.\ composite from the spiderweb galaxy template and one additional
SB template---however, requesting the template to be in agreement with all
available flux density upper limits. Furthermore, here, we required the total
template to match the median measured 1.4\,GHz flux density of the observed
IFRS of 15\,mJy.
\begin{figure}
	\centering
		\includegraphics[width=\hsize]{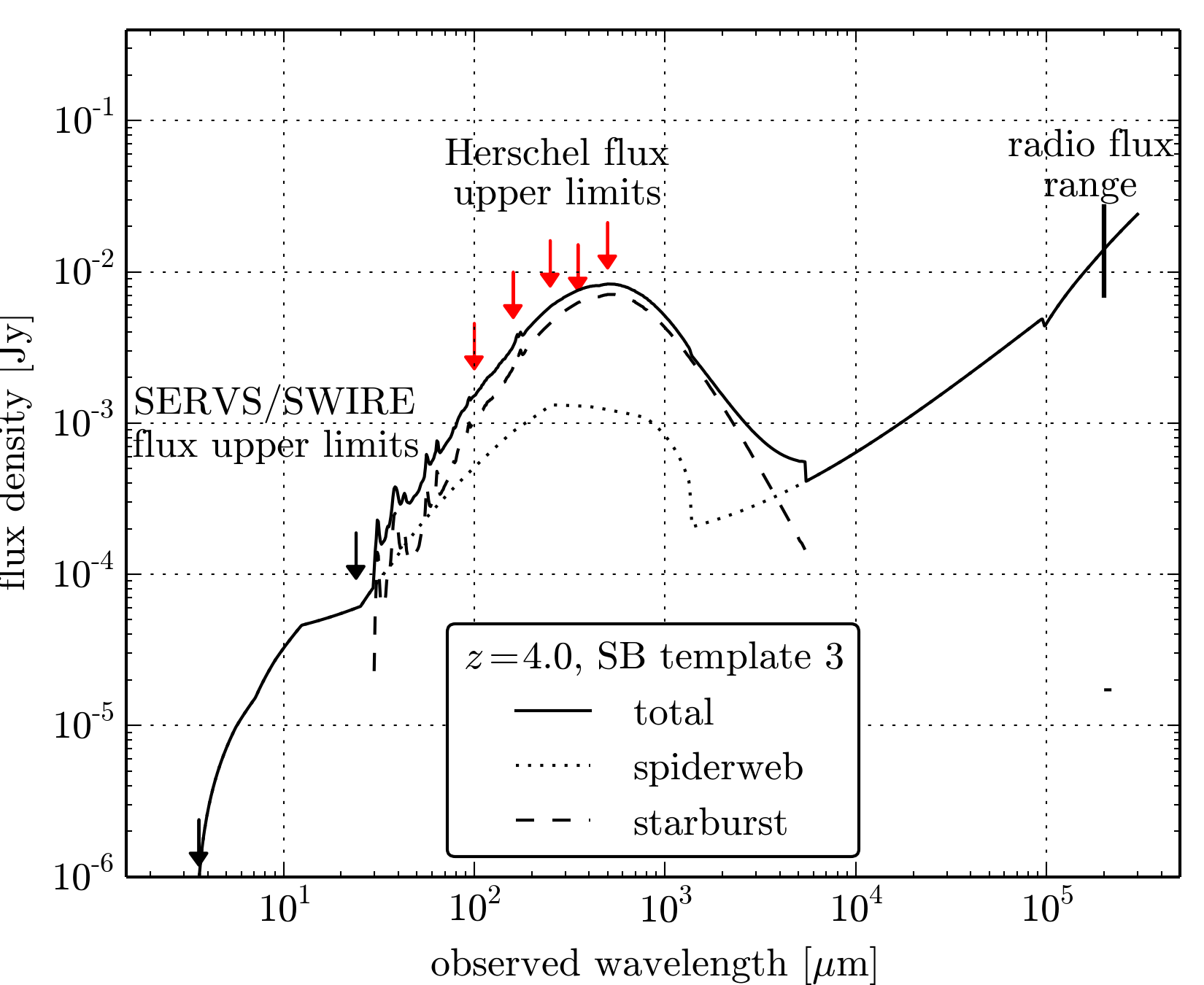}
		\caption{IR~SED~modelling for IFRS at redshift $z=4$, using the
		far-IR flux density upper limits (red arrows) measured in this work, the SERVS/SWIRE flux
		density upper limits (black arrows), and the measured radio flux
		density of 15\,mJy at 1.4\,GHz.
		The total~(solid line) emission is composed of the scaled
		spiderweb galaxy template~(dotted line) and one scaled SB
		component~(dashed line). Note that the limited wavelength coverage of the
		SB template caused the leaps at 30\,$\mu$m and 5450\,$\mu$m. The fit
		represents an upper limit on the IR~emission of IFRS and is defined by
		matching the radio flux density of 15\,mJy at 1.4\,GHz and providing the
		lowest deviation between all available flux density upper limits and the
		total~SED.
		Note that the dash at 21\,cm (flux density of $\approx 20\,\mu$Jy) is the
		contribution of the additional SB component to the 1.4\,GHz flux density.}
	\label{fig:IRSEDfit_withradio}
\end{figure}
Figure~\ref{fig:IRSEDfit_withradio} shows the resulting maximum model
for $z = 4$.\par

We found that the contribution of the additional SB component to the
radio emission, calculated based on the radio-IR correlation, is negligible compared
to the radio emission of the spiderweb galaxy template. In case of $z = 4$, this
SB contribution is around 20\,$\mu$Jy at 1.4\,GHz which is more than two
orders of magnitude lower than the contribution of the spiderweb galaxy
template at this frequency.\par

Based on the results of this modelling, we calculated the
maximum total IR luminosity by integrating the flux density of the
total SED template between 8\,$\mu$m and 1000\,$\mu$m in the rest frame.
Furthermore, we decomposed the maximum IR luminosity in the SB
contribution and the AGN contribution. \citet{Seymour2012} studied the IR properties of the spiderweb
galaxy and find that 59\% of its IR luminosity is contributed from the AGN and
41\% from the star forming activity.
We used these fractional numbers and decomposed the contribution of the
spiderweb galaxy template correspondingly.
\begin{figure}
	\centering
		\includegraphics[width=\hsize]{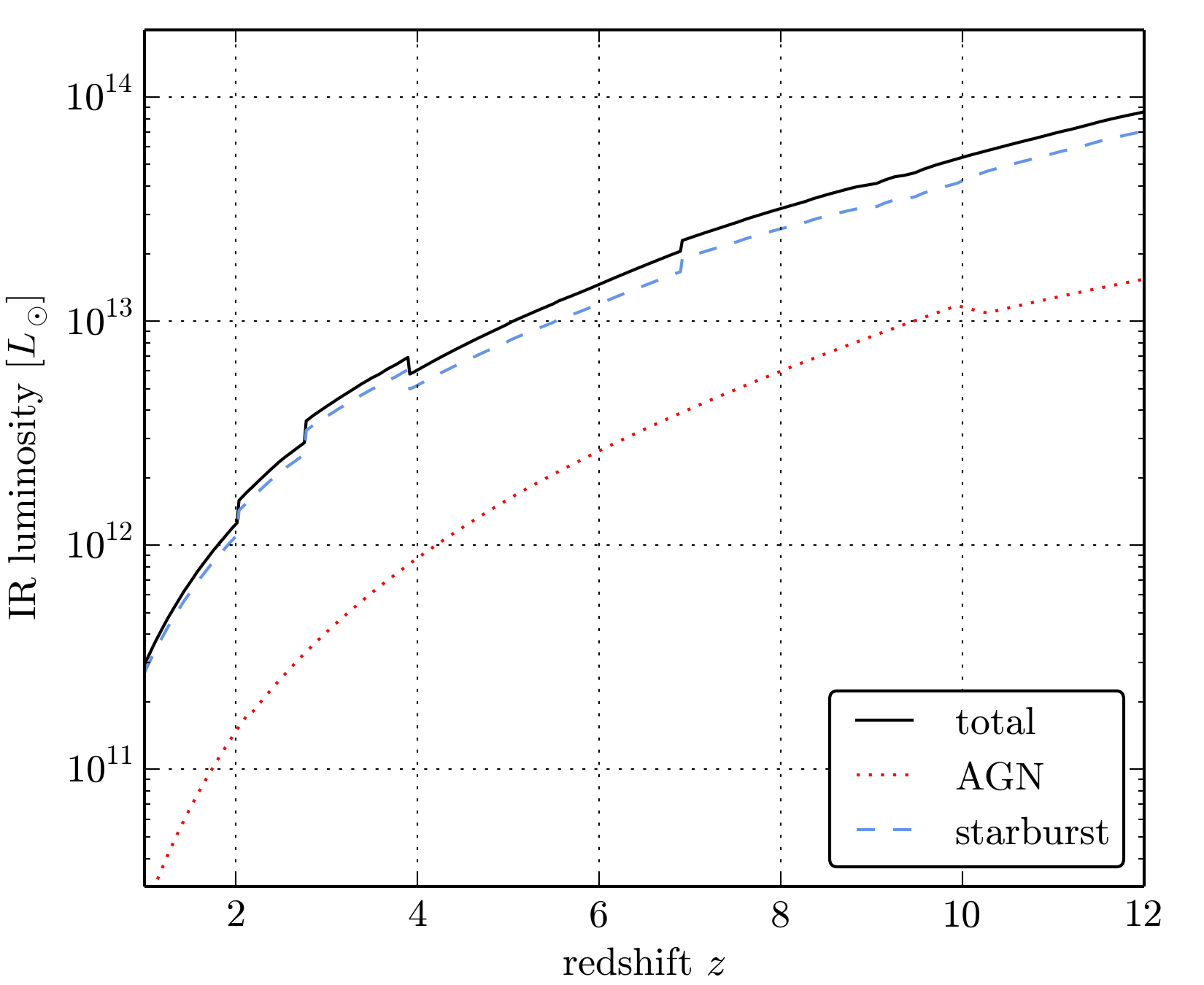}
		\caption{Maximum infrared luminosity versus redshift for the AGN~(red
		dotted line) component, the SB~(blue dashed line) component, and for the total IR
		luminosity (black solid line). The underlying model (see
		Fig.~\ref{fig:IRSEDfit_withradio} for an example) matched the measured radio
		flux density of 15\,mJy at 1.4\,GHz. We note that the total IR luminosity and the SB IR luminosity are
		upper limits.}
	\label{fig:IRluminosity-z_radio}
\end{figure}
The maximum total IR luminosity and its decomposition in AGN and SB
contribution as a function of redshift are shown in
Fig.~\ref{fig:IRluminosity-z_radio}. We found that, based on this
modelling, IFRS are constrained to IR luminosities below $10^{12}L_\odot$ at
redshifts $z\lesssim 2$. At $z \lesssim 5$, IFRS can have a maximum luminosity
of $10^{13}L_\odot$. At higher redshifts, we cannot exclude IFRS to be HyLIRGs
which are defined by IR luminosities above $10^{13}L_\odot$.\par

We note that the luminosity upper limits obtained in this approach are
up to one order of magnitude lower than those inferred in
Sect.~\ref{IR_SED_modelling_IRonly} and presented in
Fig.~\ref{fig:IRluminosity-z}. This shows that it is crucial to include the
measured radio flux density in the IR SED modelling. A
modelling just based on the IR flux density upper limits is physically
inappropriate and results in excessive IR luminosities.\par

The numbers of the maximum total IR luminosity and the SB contribution
are upper limits since they were obtained by a model aiming at the lowest deviation to
the available IR flux density upper limits. In contrast, the IR luminosities of
the AGN component can be considered to be rough estimates because of their
direct link to the measured radio flux density.
However, we point out that the IR emission of AGNs at high redshifts is
known to spread by more than one order of magnitude~\citep{Drouart2014}. It
should be emphasised that we assumed IFRS to be similar to HzRGs in this approach. Although several
indications for this similarity have been found so far, it is not proven yet.
While the luminosities shown in Fig.~\ref{fig:IRluminosity-z_radio} can be
considered to be more realistic numbers, the IR luminosity upper limits
presented in Fig.~\ref{fig:IRluminosity-z} are absolute upper limits which
should not be exceeded, independent of the nature of IFRS.\par

In our broad-band SED modelling described in Sect.~\ref{broadbandmodelling}, we
found that the local radio galaxy Cygnus~A, the local CSS source 3C\,48
and the RL quasar 3C\,273 provide appropriate templates to explain the SED characteristics of IFRS, too. As mentioned
above, these modified templates shifted to $z\geq 1$ fulfil the selection
criteria of HzRGs. Therefore, we also used the SEDs of these galaxies as basic templates,
i.e.\ containing an AGN and potentially a certain contribution from star forming
activity, instead of the spiderweb galaxy. We performed the modelling
in the same way as described above, i.e.\ using the scaled SED of
Cygnus~A, 3C\,48 and 3C\,273, respectively, as basic template and
adding the most appropriate and scaled SB template, matching all available flux density upper limits and the measured
1.4\,GHz flux density. However, in case of Cygnus~A, 3C\,48 and
3C\,273, we did not find appropriate numbers in the literature to convert the IR emission of these
objects into an AGN and an SB contribution as we did for the spiderweb galaxy
based on the results from \citet{Seymour2012}. Therefore, we could only compare
the total IR luminosity upper limits. We found that these numbers derived based
on Cygnus~A, 3C\,48 and 3C\,273 as basic templates are very similar to
the numbers based on the spiderweb galaxy template shown in Fig.~\ref{fig:IRluminosity-z_radio},
differing by not more than a factor of two.
This finding provides evidence that the total IR luminosity upper limits
presented in Fig.~\ref{fig:IRluminosity-z_radio} are indeed general upper
limits, independent on the assumed nature of IFRS.\par

\subsubsection{Estimates of star formation and black hole accretion rate upper
limits}
\label{rates}

Based on the maximum IR~luminosities measured in
Sect.~\ref{IRmodelling+radio}, we derived upper limit estimates of the
star formation rate~(SFR), causing the IR~emission given by the SB~component,
and estimates of the black hole accretion rate, generating the
AGN~contribution to the IR~flux. Since these numbers can only be considered as rough estimates, we used
very general and simple approaches. We calculated the SFR based on the
IR~luminosity~$L_\mathrm{SB}^\mathrm{IR}$ integrated between 8\,$\mu$m and
1000\,$\mu$m,
\begin{align}
\mathrm{SFR}\,[\mathrm{M}_\odot\,\mathrm{yr}^{-1}]=1.72\times 10^{-10}\times
L_\mathrm{SB}^\mathrm{IR}\,[L_\odot]~,
\end{align}
following the relation for local galaxies from \citet{Kennicutt1998}.\par

We estimated the black hole accretion rate~($\dot{M}_\mathrm{BH}^\mathrm{acc}$)
from the equation
\begin{align}
\kappa_\mathrm{AGN}^\mathrm{Bol} L_\mathrm{AGN}^\mathrm{IR} = \epsilon
\dot{M}_\mathrm{BH}^\mathrm{acc}c^2~,
\end{align}
based on the IR~luminosity~$L_\mathrm{AGN}^\mathrm{IR}$ of the AGN. This
relation assumes that a fraction of the rest-frame energy of the matter accreting onto the black
hole is converted into radiation over the whole electromagnetic spectrum. The
conversion is given by the efficiency factor~$\epsilon$, only slightly
constrained by empirical studies. While for example \citet{Shankar2010}
suggest $\epsilon>0.2$ based on quasar clustering, others like
\citet{Davis2011} find $0.06<\epsilon<0.4$, depending on the mass.
Here, we used a conservative number $\epsilon = 0.1$ since we were interested in
upper limits for the black hole accretion rate. Furthermore, we needed to make
an assumption for the factor~$\kappa$, converting the IR~AGN~luminosity to the
bolometric luminosity. This factor is uncertain and can vary between 1.4 and 15.
We adopted the number from \citet{Drouart2014}, $\kappa=6$, assuming an RL~AGN template.\par

\begin{figure}
	\centering
		\includegraphics[width=\hsize]{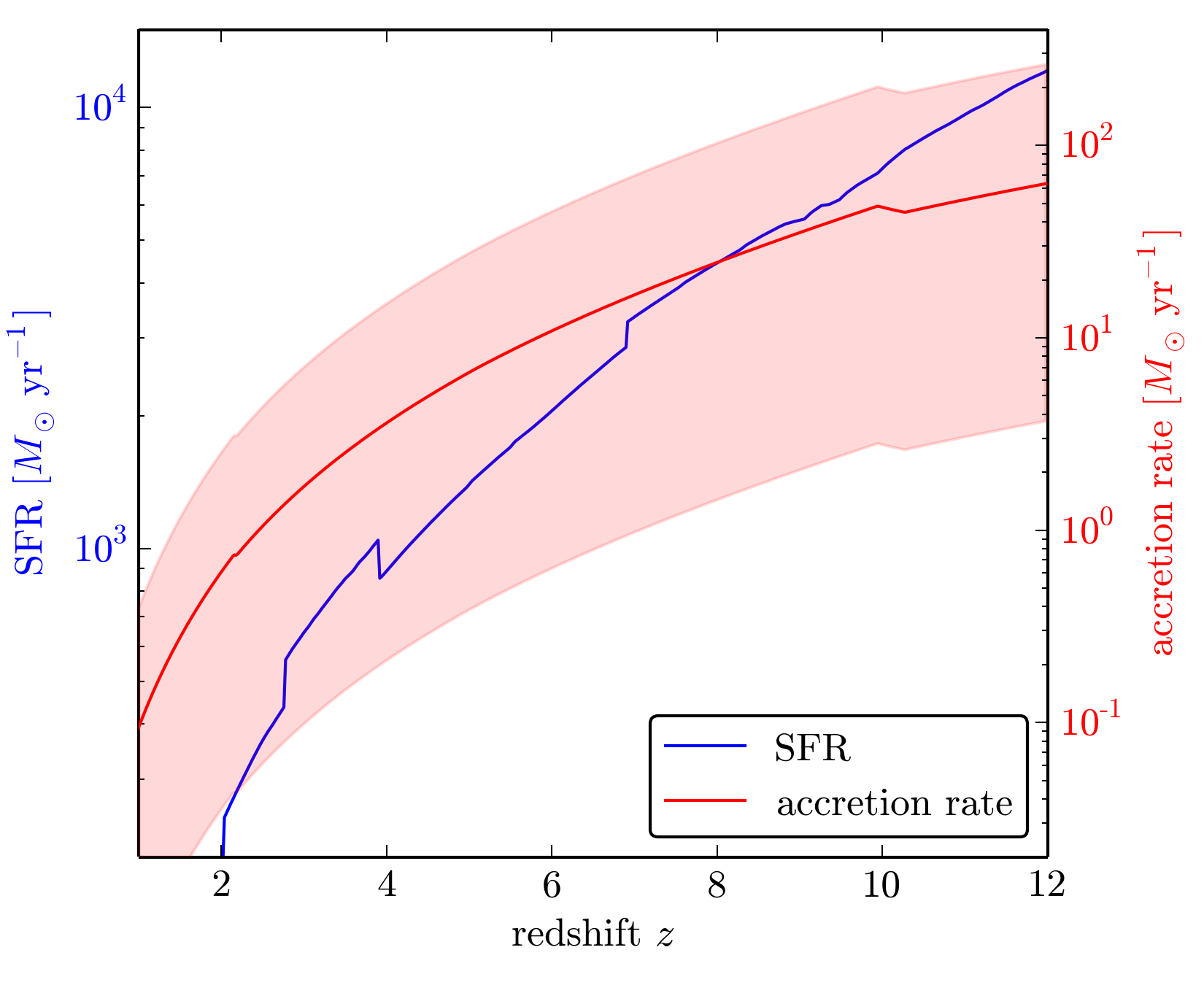}
		\caption{Star formation rate~(blue line) and accretion rate~(red line) versus
		redshift. The calculation of both rates is based on the modelling
		described in Sect.~\ref{IRmodelling+radio}, matching the measured 1.4\,GHz
		flux density of 15\,mJy and providing the lowest deviation to all available flux density upper
		limits. The red shaded area represents the uncertainty in the determination of
		the BH accretion rate, based on varied numbers for the efficiency
		factor~$\epsilon$ and the conversion factor~$\kappa$. We note that the SFR is
		meant to be an upper limit since the SB contribution is only constrained by
		upper limits.}
	\label{fig:Spiderweb_plot_rates}
\end{figure}

Figure~\ref{fig:Spiderweb_plot_rates} shows the estimated upper limit of
the SFR and an estimate of the black hole accretion rate versus
redshift based on the maximum IR luminosities derived in
Sect.~\ref{IRmodelling+radio}.
For the accretion rate, we also showed the related uncertainty, arising from the
assumptions made for the efficiency factor~$\epsilon$ and the conversion
factor~$\kappa$ as discussed above. Varying $\epsilon$ between 0.06 and 0.4, and
$\kappa$ between 1.4 and 15, representing the range of reasonable numbers,
results in the uncertainties shown by the red shaded area in
Fig.~\ref{fig:Spiderweb_plot_rates}.\par

We found that IFRS at $z\lesssim 4.5$ are constrained by SFRs of a few
hundred solar masses per year. The SFR upper limit increases with redshift and
exceeds $10^4\,M_\odot\,\mathrm{yr}^{-1}$ at $z\approx 11$. \citet{Drouart2014}
find SFRs for their HzRGs in the range of $100\,M_\odot\,\mathrm{yr}^{-1}$ to
$5000\,M_\odot\,\mathrm{yr}^{-1}$. These numbers are in agreement with our
results, considering that the SFRs calculated here are meant to be upper
limits.\par

The estimates of the black hole accretion rate are below
$1\,M_\odot\,\mathrm{yr}^{-1}$ at $z\lesssim 3$ and below $10\,M_\odot\,\mathrm{yr}^{-1}$ at $z\approx
6$.
At redshifts $z\geq 10$, the accretion rate is below a few tens of
solar masses per year. \citet{Drouart2014} find accretion rates between
$1\,M_\odot\,\mathrm{yr}^{-1}$ and $100\,M_\odot\,\mathrm{yr}^{-1}$ for their
sample of HzRGs. Again, our results are in agreement with the numbers
presented by \citeauthor{Drouart2014}. Considering that the HzRGs in the
sample of \citeauthor{Drouart2014} are at $z\lesssim 5$, IFRS seem to show
lower accretion rates if they are located at similar redshifts. In contrast to
the SFR estimated in this work, which are meant to be upper limits as discussed
above, the black hole accretion rates are considered to be rough estimates under
the assumption that the SEDs of IFRS are similar to those of HzRGs.
However, the assumption for the conversion factor~$\kappa$ and the
efficiency factor~$\epsilon$ and the scattering of the IR luminosity of
AGNs~\citep{Drouart2014} add significant uncertainties.\par

All our findings in this IR~SED modelling are in agreement with the
results by \citet{Drouart2014} for HzRGs. The redshift-dependent IR~luminosity upper
limits and the directly related SFRs and black hole accretion rates derived in
this work agree with the numbers calculated by \citeauthor{Drouart2014} for their
sources which, in contrast to IFRS, partially provide FIR detections. In our
IR SED modelling, we did not find any evidence disproving the
hypothesis that IFRS are similar to HzRGs.

\section{Conclusion}
\label{conclusions}

In this work, we presented the first FIR~data of IFRS. Six IFRS have been
observed between 100\,$\mu$m and $500\,\mu$m with the instruments PACS and SPIRE
on board the \textit{Herschel} Space Observatory. None of the observed IFRS has
been detected at any of the five FIR wavelengths down to median $3\sigma$ levels
between 4.4\,mJy at 100\,$\mu$m and 17.6\,mJy at 500\,$\mu$m. Even the stacking
of the five maps of five IFRS at each wavelength did not provide a
detection.\par

We used the FIR flux density upper limits combined with the radio detections and
the SERVS/SWIRE flux density upper limits to model the broad-band SED of IFRS.
The characteristics of IFRS can only be explained by known SEDs of HzRGs (e.g.\
the spiderweb galaxy) if these templates are shifted to $z \gtrsim 10.5$ or by
the SED of Cygnus~A at $5\lesssim z \lesssim 8.5$. All other templates, for
example RL HyLIRGs, ULIRGs with an AGN in their centre, or RL quasars fail to
reproduce the SED constraints of IFRS. We also tested whether modified templates
of known galaxies can explain the characteristics of IFRS at lower redshifts $z
\lesssim 5$. In this approach, we scaled the templates in luminosity to match
the measured flux densities at 1.4\,GHz or added extinction in the rest-frame
optical and near-IR regime if required. We found that the templates of the
spiderweb galaxy, Cygnus~A, the CSS source 3C\,48, and the RL quasar 3C\,273 can
be modified to match all photometric constraints of IFRS. However, at low
redshifts, additional obscuration by dust was needed for most SED templates to
match the near-IR faintness of IFRS.\par

Although no IFRS has been found at low redshifts $z\sim 1$, our modelling
provided appropriate modified templates for this scenario. This implies that our
modelling is physically unrealistic at low redshifts, potentially because of
scaling down the luminosity and---by this---the SMBH mass which might prevent
the object to be an RL AGN. On the other hand, if the IFRS analysed in this work
are indeed at $z\sim 1$, they would form a new subclass of IFRS with different
characteristics than the IFRS with known spectroscopic redshifts. Particularly,
we showed that SED templates of star forming galaxies, Seyfert galaxies, ULIRGs,
and HyLIRGs are inappropriate to reproduce the photometric constraints of
IFRS.\par

We modelled the maximum IR~SED of IFRS based on a set of SB templates and an
AGN~contribution and using all available flux density upper limits, aiming at
measuring the maximum IR flux of IFRS. We found that this model is unphysical at
lower redshifts because of its disagreement with the radio-IR correlation. We
found that the IFRS could also be modelled by the spiderweb galaxy template,
together with an additional SB component. The related IR luminosity upper limits
are in agreement with those of HzRGs. Using these maximum IR~luminosities, we
estimated black hole accretion rates and upper limits for the SFR. These numbers
agree with the numbers of HzRGs, too.\par

In summary, the IFRS discussed here might be (a)~objects identical to known
HzRGs, but at high redshifts ($z\gtrsim 10.5$), (b)~objects similar to Cygnus~A,
but at high redshifts $5\lesssim z \lesssim 8.5$, (c)~objects similar to known
HzRGs or Cygnus~A, but scaled down in luminosity, or (d)~objects similar to
CSS~sources or RL quasars, but modified by significant additional extinction and
scaled in luminosity which makes these templates to HzRGs, too. In any case,
objects which reproduce the characteristics of IFRS fulfil the selection
criteria of HzRGs. We estimated that IFRS contain SMBHs accreting at the rate of
$1\,M_\odot\,\mathrm{yr}^{-1}$ to $50\,M_\odot\,\mathrm{yr}^{-1}$, together with
star formation at a rate of up to several thousand solar masses per year in case
of an additional SB component.\par

\begin{acknowledgements}
We thank the anonymous referee who has helped to improve this paper
significantly. We also thank Nick~Seymour for valuable discussions about the IR
SED modelling and Elaine Sadler for suggestions about SED templates. AH
acknowledges funding from \textit{Bundesministerium f\"ur Wirtschaft und
Technologie} under the label 50\,OR\,1202. The author is responsible for the
content of this publication.
PACS has been developed by a consortium of institutes led by MPE
(Germany) and including UVIE (Austria); KU Leuven, CSL, IMEC (Belgium); CEA, LAM
      (France); MPIA (Germany); INAF-IFSI/OAA/OAP/OAT, LENS, SISSA (Italy); IAC
      (Spain). This development has been supported by the funding agencies BMVIT
      (Austria), ESA-PRODEX (Belgium), CEA/CNES (France), DLR (Germany),
      ASI/INAF (Italy), and CICYT/MCYT (Spain). SPIRE has been developed by a
      consortium of institutes led by Cardiff University (UK) and including
      Univ. Lethbridge (Canada); NAOC (China); CEA, LAM (France); IFSI, Univ.
      Padua (Italy); IAC (Spain); Stockholm Observatory (Sweden); Imperial
      College London, RAL, UCL-MSSL, UKATC, Univ. Sussex (UK); and Caltech, JPL,
      NHSC, Univ. Colorado (USA). This development has been supported by
      national funding agencies: CSA (Canada); NAOC (China); CEA, CNES, CNRS
      (France); ASI (Italy); MCINN (Spain); SNSB (Sweden); STFC, UKSA (UK); and
      NASA (USA).
      HIPE is a joint development by the \textit{Herschel} Science Ground Segment
      Consortium, consisting of ESA, the NASA \textit{Herschel} Science Center, and the
      HIFI, PACS and SPIRE consortia.
      This research has made use of the NASA/IPAC Extragalactic Database (NED)
      which is operated by the Jet Propulsion Laboratory, California Institute of
      Technology, under contract with the National Aeronautics and Space
      Administration.
\end{acknowledgements}


\bibliographystyle{aa} 
\bibliography{references} 

\end{document}